\documentclass[times,twocolumn,final]{elsarticle}
\usepackage{framed,multirow,float}

\usepackage{amssymb}
\usepackage{latexsym}

\usepackage{url}
\usepackage{xcolor}

\usepackage{hyperref}

\definecolor{newcolor}{rgb}{.8,.349,.1}

\begin{document}

%\verso{Guillon Louise \textit{et~al.}}

\begin{frontmatter}

\title{Identification of Rare Cortical Folding Patterns using Unsupervised Deep Learning}

\author[neurospin]{Louise Guillon\corref{mycorrespondingauthor}}
\cortext[mycorrespondingauthor]{Corresponding author}
\ead{louise.guillon.fondimare@gmail.com}
\author[neurospin]{Joël Chavas}
\author[adresse2]{Audrey Bénézit}
\author[adresse3]{Marie-Laure Moutard}
%\author[adresse4]{Ghislaine Dehaene}
\author[neurospin]{Denis Rivière}
\author[neurospin]{Jean-François Mangin}

\address[neurospin]{Université Paris-Saclay, CEA, CNRS, NeuroSpin, Baobab, Gif-sur-Yvette, France}
\address[adresse2]{Service de Neurologie et Réanimation Pédiatrique. Hôpital Raymond Poincaré. APHP. Garches, France}
\address[adresse3]{Service de Neuropédiatrie, Hôpital Trousseau, Hôpitaux Universitaires de l'Est Parisien, Sorbonne Université, Paris, France}
%\address[adresse4]{}

\begin{abstract}
Like fingerprints, cortical folding patterns are unique to each brain even though they follow a general species-specific organization. Some folding patterns have been linked with neurodevelopmental disorders. However, due to the high inter-individual variability, the identification of rare folding patterns that could become biomarkers remains a very complex task. This paper proposes a novel unsupervised deep learning approach to identify rare folding patterns and assess the degree of deviations that can be detected.  
To this end, we preprocess the brain MR images to focus the learning on the folding morphology and train a $\beta-VAE$ to model the inter-individual variability of the folding. We compare the detection power of the latent space and of the reconstruction errors, using synthetic benchmarks and one actual rare configuration related to the central sulcus. Finally, we assess the generalization of our method on a developmental anomaly located in another region.
Our results suggest that this method enables encoding relevant folding characteristics that can be enlightened and better interpreted based on the generative power of the $\beta-VAE$. The latent space and the reconstruction errors bring complementary information and enable the identification of rare patterns of different nature. This method generalizes well to a different region on another dataset. Code is available at \url{https://github.com/neurospin-projects/2022_lguillon_rare_folding_detection}.

\end{abstract}


\end{frontmatter}

%\linenumbers

\section{Introduction}
\begin{figure*}[ht]
    \centering
    \includegraphics[scale=0.23]{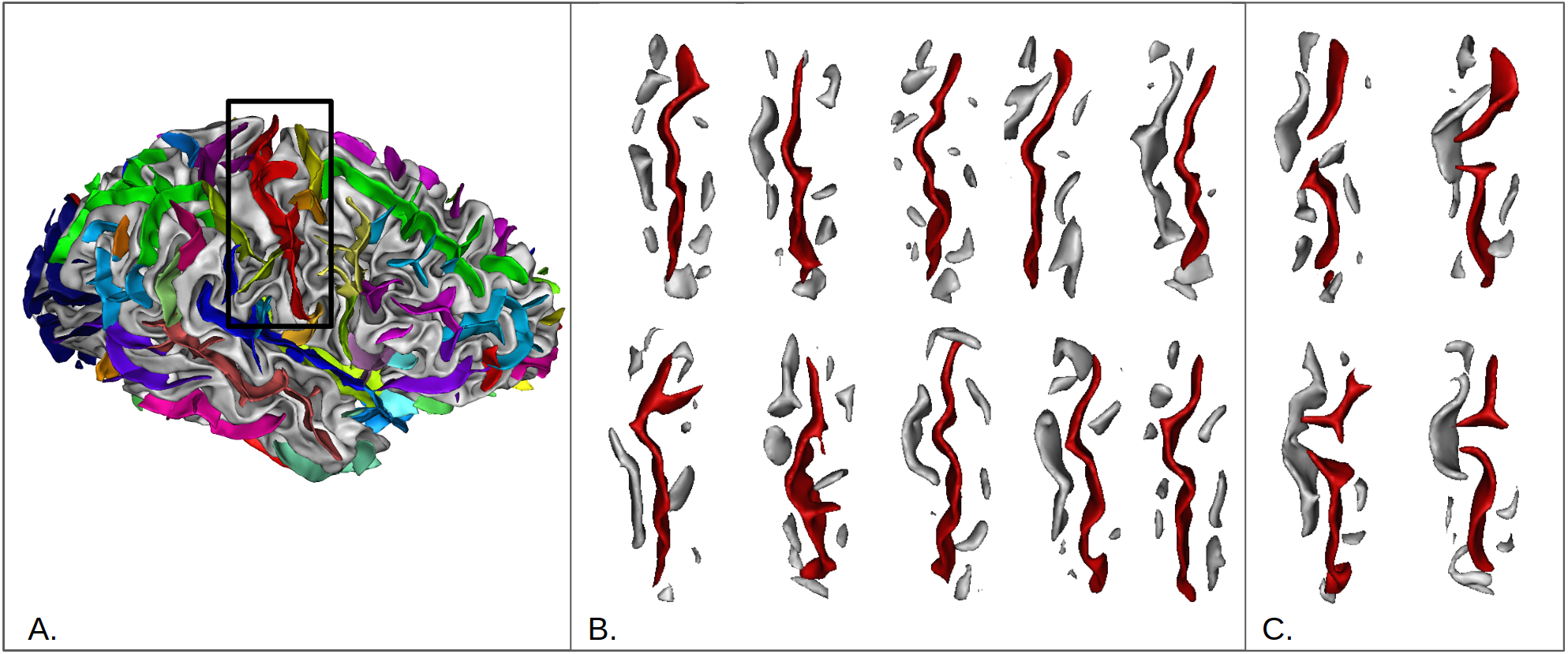}
    \caption{\textit{Central sulcus region variability.} A. Localization of the studied region of interest (ROI) on a 3D view of one right hemisphere. The colored ribbons represent sulci, defined as a negative cast of the furrows. The central sulcus is red. B. Examples of non-interrupted central sulci. C. Examples of interrupted central sulci.}
    \label{fig:cs_variability}
\end{figure*}

During gestation, the human cortex folds and gets its convoluted shape composed of gyri---the ridges of white matter---that are delimited by furrows---
the sulci. Historically, their shape and characteristics have been described by neuroanatomists based on specimens. 
In the human population, stability of the folding patterns is observed with an overall similarity of location, shape and arrangements \citep{ono_atlas_1990}. This stability is important enough to enable to define a road map and a nomenclature of sulci and to develop methods that automate sulci recognition \citep{riviere_automatic_2002,borne_automatic_2020}. Despite this homogeneity, each brain displays a unique cortical folding, acting as a fingerprint \citep{wachinger_brainprint_2015}. Fig. \ref{fig:cs_variability}A and B. show examples of the variability in the central sulcus region, which is one of the most stable. The folding variability is so complex that it has long been overlooked.

However, thanks to advances in the neuroimaging field, studies have tried to characterize sulci with elementary shapes that amount to building blocks of alternative patterns. For instance, the central sulcus is typically composed of one or several knobs \citep{yousry_localization_1997}. Similarly, the mid-fusiform sulcus presents an omega pattern \citep{weiner_mid-fusiform_2014}. In contrast, some very rare patterns have also been described, such as the interruption of the central sulcus that can be found in only about 1\% of the population \citep{mangin_plis_2019}. Four examples of interrupted central sulci in the right hemisphere are presented in Fig. \ref{fig:cs_variability}C. 

Folding patterns have also proved to be very interesting as they are related to function. For example, the central sulcus divides the cortex into the motor and the sensory areas, and specific parts of the central sulcus have been correlated with the cortical areas of the tongue, foot, and hand among others \citep{mangin_plis_2019, germann_tight_2020}. Specifically, the central sulcus main knob has been linked to the hand motricity and is called the "hand knob" \citep{yousry_localization_1997}. In another region, cingulate sulcus patterns have been associated with the inhibitory control \citep{borst_folding_2014}. 

Specific patterns were also correlated to neurodevelopmental disorders. The Power Button Sign (PBS), a rare configuration of the precentral sulcus, may be associated with a certain type of epilepsy \citep{mellerio_power_2014}. Patterns in the superior temporal sulcus (STS), central, intraparietal and frontal regions could be related to autism \citep{levitt_cortical_2003, auzias_atypical_2014,hotier_social_2017}. 
Hence, deciphering sulcal complexity and having a better understanding of the underlying shape variability is of great interest: folding patterns could become biomarkers of neurodevelopmental disorders.

Folding patterns can be analyzed with two approaches. On one hand, morphometric features can be extracted such as the depth, the surface curvature, or the opening of each sulcus. On the other hand, one can look directly at the shapes of the sulci. Working on shapes rather than on morphometric values is particularly interesting as they constitute "trait features" opposite to "state features" \citep{cachia_longitudinal_2016}. Unlike state features that can evolve during the lifespan, trait features remain fixed afterbirth. For example, the sulcal opening is a state feature because it increases with aging \citep{kochunov_age-related_2005, jin_relationship_2018}. In return, the pattern of the cingulate sulcus area is a trait feature because it is stable throughout life after infancy. This difference between trait and state features has also been demonstrated in the study of the effects of handedness on the central sulcus shape. For example, forced dextral subjects show similarities to sinistral subjects in shape, but changes in elongation mimicking dextrals, occur when they are constrained to use the right hand for writing \citep{sun_effect_2012}. Different strategies can be adopted for exploring the shapes: a finite number of shapes can be considered, using clustering for instance \citep{meng_discovering_2018, duan_exploring_2019}, or shapes can be represented in a continuous way, such as manifold-based analyses \citep{sun_effect_2012}, \citep{de_vareilles_shape_2022}. For both strategies, a first step is required to represent the folding patterns. Folding shapes' complexity can be reduced based on the similarity between different sulci \citep{sun_constructing_2009, sun_effect_2012, de_vareilles_shape_2022} or between sulcal graphs \citep{im_quantitative_2011,meng_discovering_2018}. These similarity measures are then either directly analyzed, or projected to a lower dimensional space.
 
Advances in machine learning are now opening up new possibilities for studying folding patterns, identifying typical or rare patterns, and hopefully, emerging sulcal biomarkers. Recently, a neural network classifier was used to map geometric shapes to the broken-H shape pattern in the orbitofrontal region \citep{roy_pipeline_2020}. However, this method requires having pre-identified the geometric shape to be mapped, which makes it difficult for unknown patterns to emerge. To tackle this issue, unsupervised deep learning techniques seem to be promising. Thus, two unsupervised deep learning models, a $\beta-VAE$ and SimCLR, were compared in the task of identifying typical patterns in the cingulate region \citep{guillon_unsupervised_2022}. Other works have focused on the task of identifying abnormal folding patterns thanks to unsupervised deep learning in the region of the superior temporal sulcus branches \citep{guillon_detection_2021}. Anomaly detection has been a subject of great interest in the domain of biomedical imaging: many studies, including for brain MR images, have tried to identify abnormal samples \citep{chalapathy_deep_2019, fernando_deep_2022}. A common framework is to use auto-encoders as they implement a latent space with fewer dimensions than the input which makes it hard to encode uncommon features. Then, the identification of anomalies is usually performed based on the reconstruction error rather than in the latent space.

In this work, we investigate whether an unsupervised deep learning model can learn normal folding variability to identify deviating regional patterns; and if so, what granularity of deviations can be detected? 
%Therefore, we seek to develop a method to automatically detect rare or abnormal folding patterns based on unsupervised deep learning. 
Here, we define granularity as the characteristics and properties of the anomalies, such as their size or nature. The analysis of the granularity that can be identified aims to characterize the abnormal features that can be detected and at what level of detail. We also seek to describe which space is the most relevant to identify deviating patterns: is it based on the reconstruction error, in the input space, that is to say in our case, the \textit{folding space}, or is it the latent space?

Folding mechanisms may lead to both global and regional anomalies and these two scales have led to correlations with function disorders \citep{fernandez_cerebral_2016}. Here, we focus on regional patterns rather than on a global representation.

Specifically, we concentrate on the central sulcus which is a good candidate for our work. Indeed, it is one of the first folds to appear and it is stable enough to be a first step in modeling inter-individual variability. More importantly, usually long and continuous, the central sulcus can be interrupted in very rare cases, making interrupted central sulci relevant patterns to assess our method. Finally, this region is of clinical interest as it is linked to hand motricity and asymmetries have been described \citep{sun_effect_2012, bo_asymmetries_2015}.

To perform our study, we worked on the HCP database \citep{van_essen_wu-minn_2013}. From the MR images, we focused on the folding morphology of the central sulcus area with a specific preprocessing that does not require labeling the sulci of the studied subjects. Indeed, in the future, we wish to be able to apply our methodology to new databases whether the sulci are labeled or not. Even if automatic recognition tools exist and are efficient, some errors may remain and thus lead to the selection of another sulcus and to a contaminated learning set, especially regarding unusual patterns. We then trained a $\beta-VAE$ to learn the inter-individual variability. Due to the small number of interrupted central sulci and to be able to characterize the detected granularity, we designed synthetic benchmarks of rare patterns to assess our methodology more reliably. We then investigated the detection power of our methodology both on the latent space and on the \textit{folding space}, using either our synthetic outliers or actual interrupted central sulci, an actual rare pattern. Finally, we assessed the generalization of our approach on another dataset presenting abnormal folding patterns in the cingulate region.

We stress out that rare patterns are not necessarily abnormal and associated with some disorders. However, whether they have a link with pathologies or not, rare patterns are interesting objects to study as they can constitute traces of neurodevelopmental processes. Therefore, in this article, the only \textit{abnormal} pattern that we study is the one in the cingulate region. We consider our synthetic benchmarks and interrupted central sulci as rare patterns.

\section{Material and methods}

\subsection{Database}
We used T1 weighted MR images of the Human Connectome Project (HCP) dataset \citep{van_essen_wu-minn_2013}. Data were acquired on a single Siemens Skyra Connectom scanner at an isotropic resolution of 0.7mm. Subjects are healthy controls from 22 to 36 years old. In the context of our study we considered only the right hemispheres of the right-handed subjects leading to a total of 1001 subjects. \\
The long-term goal of this work is to identify rare folding patterns that have not been characterized yet. However, we first need to assess our method. To do so, we decided to work on a rare pattern already described, the interrupted central sulci (CS). A previous study identified in this database seven sulci in the right hemisphere and two in the left \citep{mangin_plis_2019}. The identification was based on the depth profiles of the sulcal pit maps, which are defined as the locally deepest point of the cortical surface \citep{lohmann_deep_2008}, and it was then visually confirmed. We chose to work on the right hemisphere rather than on the left in order to have the highest number of rare patterns.

\subsection{Folding representation}
\begin{figure*}
    \centering
    \includegraphics[scale=0.25]{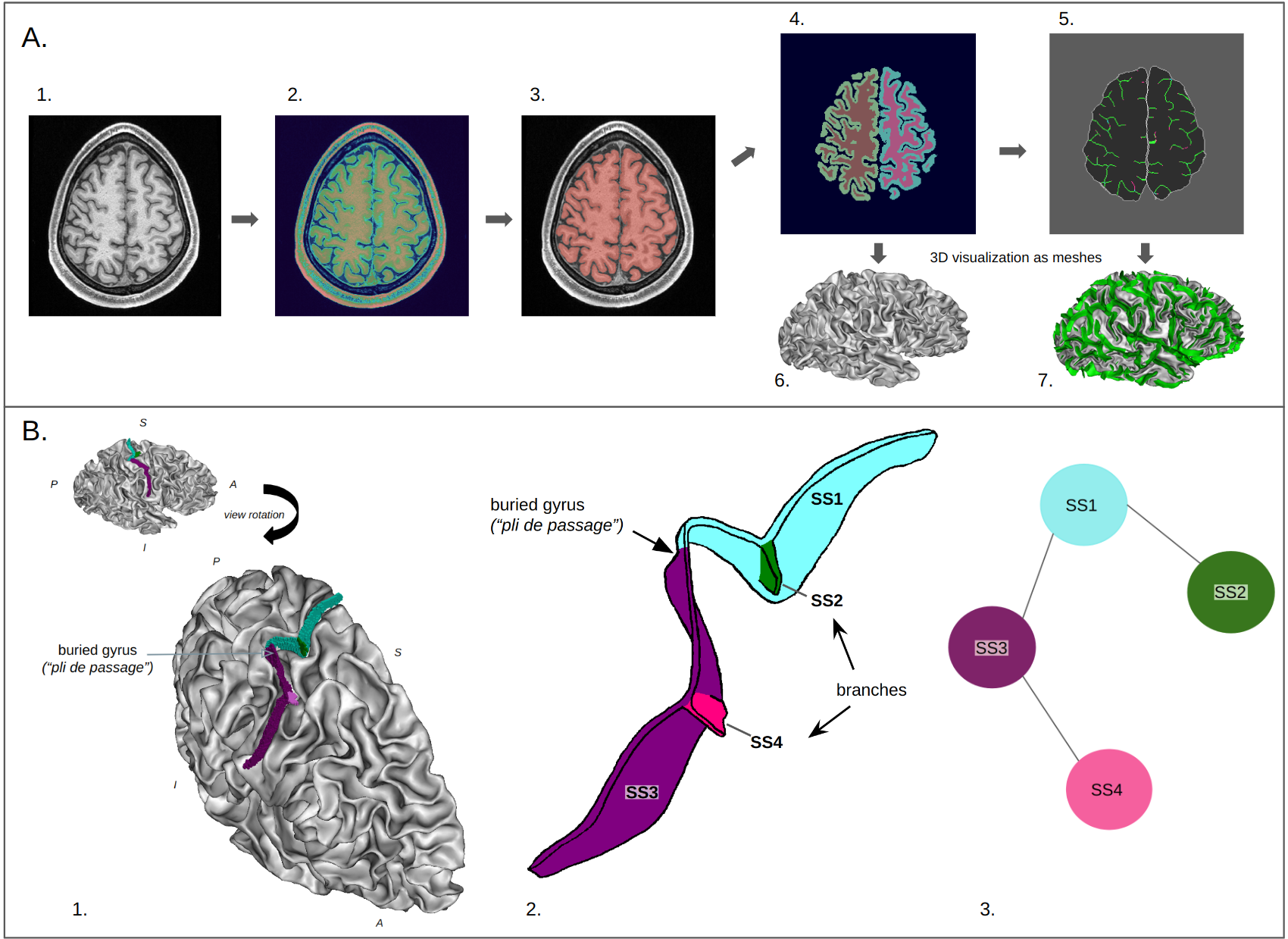}
    \caption{\textit{Overview of the BrainVISA/Morphologist pipeline's main steps and of the folds representation.} \textit{A. Main steps of BrainVISA/Morphologist pipeline.} 1. Raw T1-w MRI, 2.  Bias-corrected image, 3. Segmentation of the brain, 4. Segmentation of the hemispheres and of the grey and white matter, 5. Skeleton representation of the folding graph, representing a negative cast of the 4. 6. Mesh representation of the white matter of the right hemisphere, 7. Folding graph that represents the folds (in green) as the negative cast of the white matter of the right hemisphere (white mesh). \textit{B. Folds representation.} 1. Example of a central sulcus, which is composed of several elementary entities called simple surfaces (SS). (Orientation: A: Anterior, P: Posterior, S: Superior, I: Inferior). 2. Corresponding schematic representation of the sulcus represented in 1, which is formed by four simple surfaces. Depth variation caused by the buried gyrus and the presence of two branches lead to the division into four different simple surfaces. 3. Corresponding folding graph.}
    \label{fig:morphologist}
\end{figure*}

In this work, we consider the folds as the skeleton of a negative cast of the brain, that is to say, voxels located in the cerebrospinal fluid, (Fig.\ref{fig:morphologist}A.5) \citep{mangin_3d_1995}, which can be represented as ribbons located between the gyri (green ribbons in Fig.\ref{fig:morphologist}A.7). Folding or sulcal patterns are defined as the combination and arrangements of shapes of one or several folds.

\begin{figure*}[ht]
    \centering
    \includegraphics[scale=0.3]{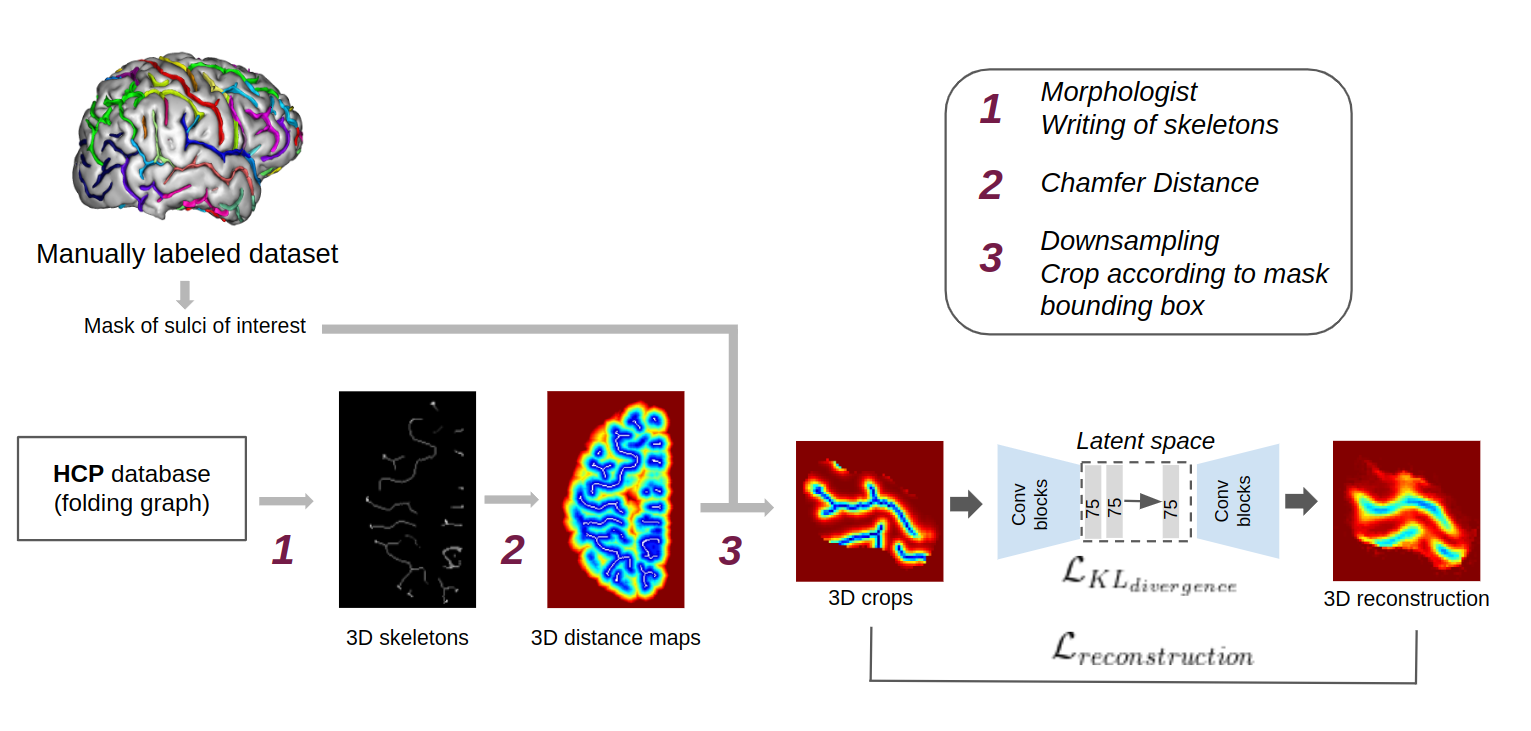}
    \caption{\textit{Pipeline.} A mask of the central sulcus area is defined based on a distinct manually labeled dataset. HCP is processed with Morphologist to obtain folding graphs, which are used to obtain 3D images of skeletons (1). The Chamfer distance is applied to the skeletons to obtain geodesic distance maps (2). Distance maps are then downsampled and cropped according to the mask (3) and are fed as input to a $\beta-VAE$.}
    \label{fig:pipeline}
\end{figure*}

\paragraph{BrainVisa/Morphologist pipeline}
Structural MR images hold numerous pieces of information beyond the morphology of cortical folding. In order to focus on the folding characteristics, we developed a preprocessing pipeline.
The raw MR images are first processed by the BrainVISA/Morphologist software (\url{https://brainvisa.info/}). This pipeline is composed of several steps that include skull stripping, bias correction, segmentation of the brain and of the hemispheres, skeletonization of the grey matter and the cerebrospinal fluid union (Fig.\ref{fig:morphologist}A) \citep{riviere_automatic_2002}. This step leads to so-called skeletons, 3D images representing only the folding which is then segmented into simple surfaces (SS) depending on various parameters such as the sulcal depth or topological properties (Fig.\ref{fig:morphologist}B) \citep{mangin_3d_1995}. For example, in Fig.\ref{fig:morphologist}B, small branches (SS2 and SS4) are represented as different simple surfaces from the main ones, SS1 and SS3. The depth variation resulting from the buried gyrus leads to two distinct simple surfaces (SS1 and SS3). Therefore, in this case, the central sulcus is composed of four simple surfaces. All in all, the obtained outputs are 3D images that correspond to a negative cast of the brain (first step of Fig.\ref{fig:pipeline}).

\paragraph{From skeletons to distance maps}\label{distmap_explanation}
Skeleton-based images have proved to be relevant and were the object of previous studies for fold recognition \citep{borne_automatic_2020, borne_automatic_2021}, as well as for folding patterns representation \citep{guillon_detection_2021, guillon_unsupervised_2022}. However, in this work, we applied an additional preprocessing step to convert our skeleton images into distance maps. Indeed, we believe that skeletons have several shortcomings that could limit the performance of our approach. 
First, skeletons are binary images representing the folding patterns; hence, they are very sparse images. Therefore, only very few voxels hold explicit sulcal information in skeletons. As shown in Fig.\ref{fig:dstrb_deletion}, on average, sulci represent 3500 voxels in our region of interest (ROI), which stand for less than 5\% of the voxels. We argue that it would be interesting to have a representation where the information devoted to folding patterns is more distributed. In addition, skeleton images are not smooth and the local regions of interest in our application correspond to high frequencies making the skeleton details harder to represent and reconstruct. It is also complex to reconstruct folding patterns in skeletons as they are not continuous images, so there is no notion of the proximity of a voxel to a sulcus. This makes the reconstruction error and the gradient-based learning less efficient. Finally, distance maps are built based on the whole hemisphere; therefore, if we work on a ROI, they give, especially near the border of the ROI, information about objects outside the ROI, which is not the case for crops based on skeletons.
Therefore, to tackle these shortcomings, we convert the resulting skeletons into distance maps based on the Chamfer distance, which approximates the Euclidean distance. Sulci are considered objects, and the further away a sulcus is, the larger the value of a voxel is. An example of a distance map is presented in Fig.\ref{fig:pipeline} step 2.

\paragraph{Focusing on a single region: crop definition}
As we are interested in capturing the local folding patterns variability, such as the hand knob, rather than the global hemisphere-wide arrangements of sulci, we chose to focus our study on a sub-region of the right hemisphere, the central sulcus area.
To define the ROI, we learned a mask of the central sulcus over a manually labeled dataset comprising 62 healthy controls \citep{borne_automatic_2020}. Subjects are first affinely registered to the ICBMc2009 space and resampled to an isotropic resolution of 1 mm. Then, for each subject, a mask is incremented for all the voxels of the central sulcus represented as a set of simple surfaces. The resulting mask is slightly dilated by 5 mm to include potential central sulcus locations not represented in our database. 
We crop the distance maps of the HCP subjects according to the mask bounding box using the same affine normalization procedure (step 3 of Fig.\ref{fig:pipeline}). The mask is applied on the fly during the training of our network. 

\paragraph{Distance maps and folds visualization}
Data visualization, and shape characterization in particular, can be performed directly based on the distance maps. However, this type of input enables only 2D slice views. To better visualize the folds of our crops, we binarize our distance maps with an empirically defined threshold of 0.4 and convert them to meshes.

\subsection{Learning a Representation of the Folding Variability}
\subsubsection{beta-VAE}
In order to identify rare patterns we first seek to model the inter-individual variability. In the outlier detection field based on unsupervised methods, auto-encoder (AE) models are widely used as they implement a latent space, also known as the bottleneck, that has far fewer dimensions than the input space. Usually, training is performed only on control subjects. The assumption is that the model learns a representation of the \textit{normal} variability and that at inference when facing outliers, it will not be able to encode and reconstruct them as well as control data. 
%During inference, both control and outlier data are encoded. The assumption is that the model will not be able to reconstruct outliers as well as control data.
To overcome some shortcomings of simple convolutional AE and to regularize the latent space, variational auto-encoder (VAE) was introduced \citep{kingma_auto-encoding_2014}. Its strength also lies in its generative power, enabling not only to reconstruct but also to generate new data. Other AE-based models have been used for anomaly detection in the biomedical field such as Generative Adversarial Networks (GAN) \citep{schlegl_unsupervised_2017, schlegl_f-anogan_2019}, which were then transposed to brain images \citep{simarro_viana_unsupervised_2021}. A comparison of AE models showed that VAE was one of the most efficient in brain MR images \citep{baur_autoencoders_2020}. In the context of representation learning of folding patterns, VAE was proved to be well adapted \citep{guillon_detection_2021, guillon_unsupervised_2022}.

In the VAE framework, a sample of input space $\mathcal{X}$ is mapped to a distribution in a latent space $\mathcal{Z}$ of $L$ dimensions, by an encoder $\theta$. A vector z is then drawn from this distribution and reconstructed by a decoder $\phi$. The objective function seeks to minimize both the reconstruction error and the Kullback-Liebler (KL) divergence ($\mathcal{D}_{KL}$). The model is thus trained to maximize:

\begin{equation}
\label{loss}
    \mathcal{L}(\theta,\phi;\textbf{x},\textbf{z},\beta)=\mathbb{E}_{q_{\phi}(\textbf{z}|\textbf{x})}[\textup{log}p_{\theta}(\textbf{x}|\textbf{z})]-\beta\mathcal{D}_{KL}(q_{\phi}(\textbf{z}|\textbf{x})||p(\textbf{z}))
\end{equation}

\noindent where $p(\textbf{z})$ refers to the prior distribution (in this work, a reduced centered Gaussian distribution) which is approximated with $q_{\phi}(\textbf{z}|\textbf{x})$, the posterior distribution.
$\beta-VAE$ is an extension of the VAE where the KL divergence is weighted by $\beta$ \citep{higgins_beta-vae_2017}. 
 
\subsubsection{Training procedure}
\paragraph{Preprocessing}
The input data of the model are the previously defined just cropped, then masked distance maps. For augmentation purposes, random rotations between [-10°, 10°], centered on the mask center, are drawn from a uniform distribution at each epoch and applied to the whole brain, before applying the mask that strictly remains at the same position. Such rotations are also sought to limit the edge effects. More precisely, the central sulcus is surrounded by two main folds, the precentral and the postcentral sulci. Parts of these sulci are included in the ROI. Therefore, rotating the distance map \textit{under} the mask enables to capture a wider context and to try to limit their influence. 
We observe that skeleton voxels equal 0 in the initial distance maps $X$ and the values increase with the distance to a sulcus, possibly ranging up to 10 mm. Potential reconstruction errors near the sulci would be minor compared to the voxels located far from them at the edge of the mask, whereas we wish the model to concentrate more on the sulci. Thus, to limit the impact of distance variability far from the sulci, we perform a normalization according to equation \ref{normalization}, resulting in values between [0, 1], with the highest values on the folds and a saturation at about 4-5 mm which corresponds to half of the typical width of gyri.  An example is shown in Annex 1 of the supplementary. Finally, we apply a small padding, resulting in samples of dimensions 80 x 80 x 96.

\begin{equation}
\label{normalization}
X_{norm}=1-[2\frac{1}{1+e^{-X}}-1]
\end{equation}

\paragraph{Training}\label{training}
Dataset was split into train, validation, and test sets of respectively 640, 161, and 200 subjects. Training is only performed on control data, all identified interrupted central sulci (CS) were added to the test set. The interrupted central sulci were identified based on the detection of the two main sulcal pits of the central sulcus, between which a depth profile was computed to determine the depth of the "pli de passage frontal-moyen" (PPFM), usually a buried gyrus in the sulcus. Subjects with a shallow PPFM were then manually inspected to determine whether the surrounding central sulci were interrupted \citep{mangin_plis_2019}. However, we point out that there may remain some undetected interrupted central sulci in the training set as all subjects were not individually inspected. To model the normal inter-individual variability, we used a classic convolutional $\beta-VAE$ of depth 3. In order to choose the best values for $\beta$ and latent space dimension L, we performed a gridsearch ($\beta$=2-10, L=4-150). To assess each parameter configuration, we used two criteria. Our first criterion is the reconstruction quality. Indeed, we seek to leverage the reconstruction and generative power of the $\beta-VAE$, hence the reconstruction quality must be sufficient. Our second criterion is the detection power on a proxy for the interrupted central sulci. The pre-central and post-central sulci demonstrate some similarities with the central sulcus in terms of orientation, size, and shape. However, they tend to be more interrupted and to present a higher number of ramifications. Therefore, we used the HCP dataset crops of these two other regions as fake outliers. We selected only pre- and post-central sulci which presented some ambiguities with the central sulcus based on the procedure described in Annex 2 of the supplementary. Finally, our ambiguous set was composed of 28 precentral sulci and 18 postcentral sulci. For each hyperparameter combination, we trained a $\beta-VAE$ on the train set, then a linear SVM was trained to classify between the latent codes of the validation samples and of the pre- or post-central sulci. We kept the hyperparameters that led to the best classification results and good reconstructions (based on reconstruction error and visual inspection).

\subsection{Generating synthetic rare patterns}
One of the challenges of our work is the lack of consensual rare patterns to evaluate our methodology. 
In addition, it would be interesting to be able to quantify the degree of deviation that our model is able to detect. Therefore, several sets of synthetic rare patterns were generated to be used as benchmarks. Both benchmarks were generated from the test set subjects.

\subsubsection{Deletion benchmark}
\begin{figure}[ht]
    \centering
    \includegraphics[scale=0.35]{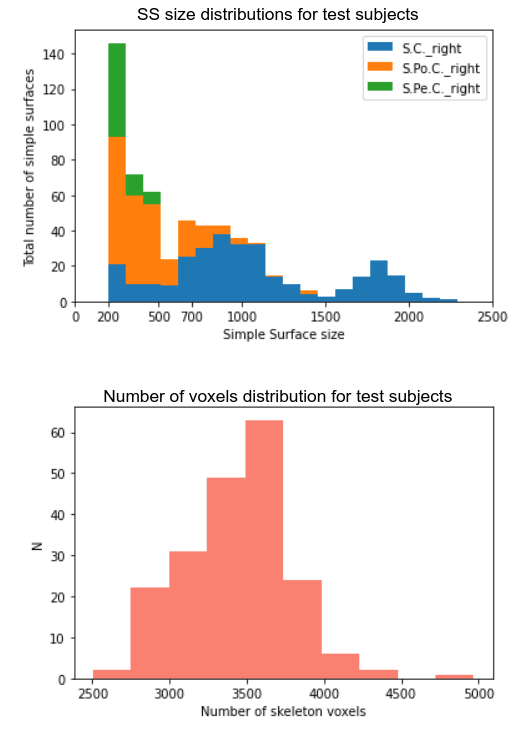}
    \caption{\textit{Skeleton's description of the test set.} Left: Stacked histogram representing the distribution of simple surfaces sizes for the test subjects for the three main sulci of our crop, the central sulcus (S.C.\_right), the precentral sulcus (S.Pe.C.\_right) and the postcentral sulcus (S.Po.C.\_right). (Note: The labeling used is automatic and therefore not entirely reliable, but these labels are sufficient to draw conclusions regarding the SS size distribution.) Right: Distribution of the number of skeletons' fold voxels for the test subjects when the mask is applied to the crops.}
    \label{fig:dstrb_deletion}
\end{figure}

\begin{figure*}[ht]
    \centering
    \includegraphics[scale=0.25]{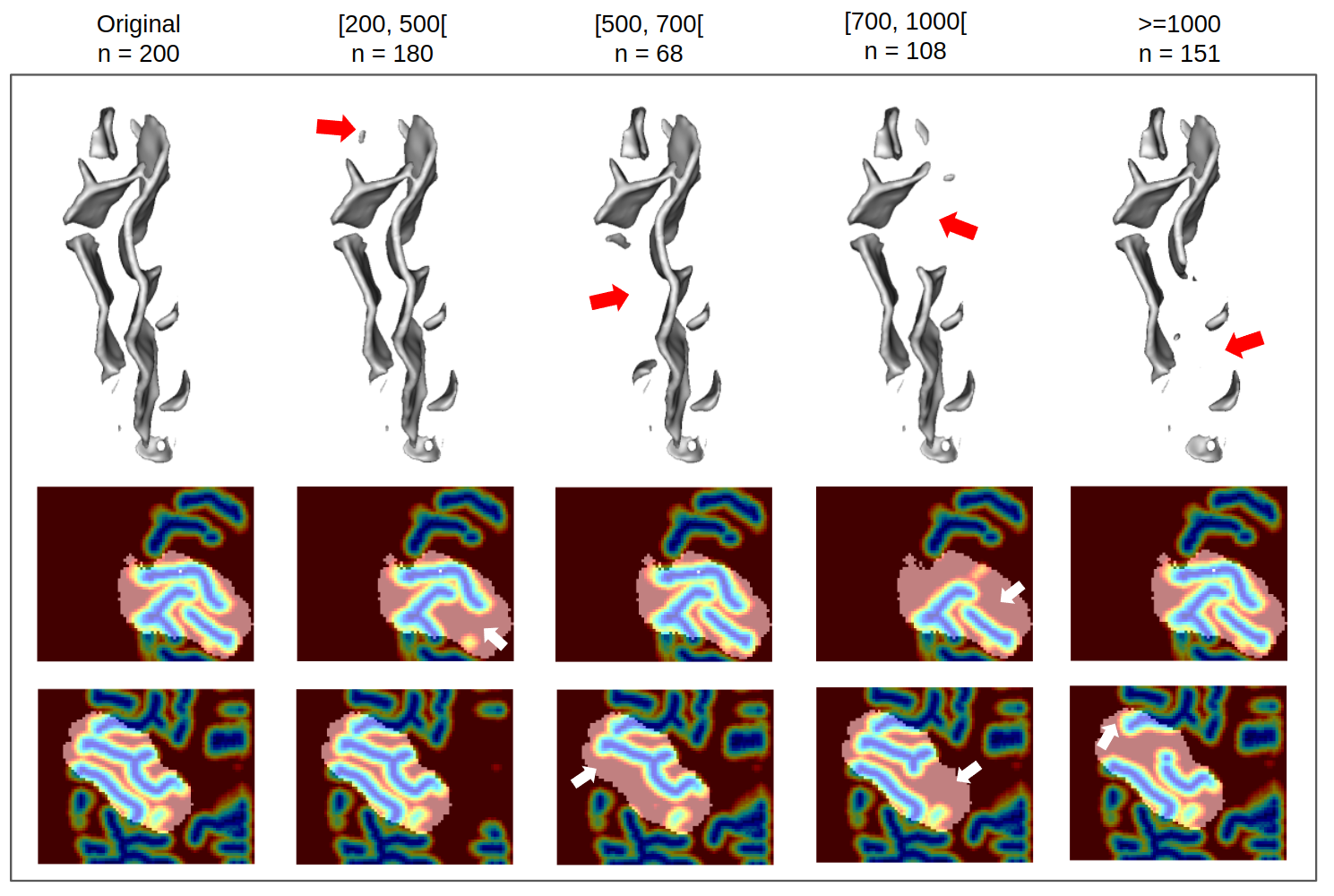}
    \caption{\textit{Deletion benchmarks.} Visualization of original sulcal pattern and its altered versions  from the four deletion benchmarks showing patterns with increasing simple surface size deleted. Upper row: Mesh visualization. Middle and bottom rows: distance maps on axial view, visualization at depths 15 and 37.}
    \label{fig:description_deletion}
\end{figure*}

Our first benchmark consists of subjects for whom we have erased one simple surface (SS). 
Erasing small simple surfaces could be a good proxy to simulate rare patterns because some fold branches may be missing in some people, or a sulcus may be shorter or absent. Large simple surfaces are less likely to be missing but allow us to assess the degree of deviation that can be detected. \\
Deleting simple surfaces directly on the distance maps would not be interesting as the voxels next to the simple surface indicate the SS position. To tackle this issue, the suppression was done during the generation of the raw skeletons. The distance map is then computed based on the pruned skeletons.
To analyze the granularity of anomaly that can be detected by our method, we generated several benchmarks which vary according to the size of the deleted simple surface (SS). As such, we created four sets where SS size was between 200-500 voxels, 500-700 voxels, 700-1000 voxels, and simple surfaces of more than 1000 voxels. In the following, we name each set with the minimum number of voxels: for instance, \textit{200} corresponds to the benchmark where simple surfaces of size between 200 and 500 were erased. 
To be deleted, simple surfaces must have a number of voxels included inside the mask corresponding to the range of the different sets. If several simple surfaces meet the criteria, one is randomly chosen to be erased. Otherwise, a subject may not have a simple surface satisfying the requirements. In such cases, the subject is not included in the benchmarks. Finally, from the 200 test subjects, benchmark 200 contains 180 subjects; benchmark 500, 68; benchmark 700, 108 and benchmark 1000, 151 subjects. 
To have a better representation of the amount of deleted sulci, Fig.\ref{fig:dstrb_deletion} shows the simple surface sizes distribution in the central sulcus region. The figure shows that our crops contain a large majority of very small simple surfaces (less than 500 voxels) and far fewer large simple surfaces. The smaller simple surfaces are mostly part of the precentral and postcentral sulci, representing more than 85\% of the surfaces between 200 and 500 voxels. On the contrary, larger simple surfaces correspond to the central sulcus. Therefore, beyond deleting simple surfaces of varying sizes, the nature of the sulci and thus the location, are also different, especially between the set 200 and the others. 
The right part of Fig.\ref{fig:dstrb_deletion} shows the number of voxels corresponding to skeletons in our crops. It demonstrates the progressive intensity of anomalies when deleting simple surfaces from 200 voxels to more than 1000 voxels. Indeed, when simple surfaces of more than 1000 voxels are deleted, it corresponds to a third or a quarter of the skeleton crop.
Distance maps are then generated according to \ref{distmap_explanation}. An example is presented in Fig.\ref{fig:description_deletion}.

\subsubsection{Asymmetry benchmark}
\begin{figure}[ht]
    \centering
    \includegraphics[scale=0.2]{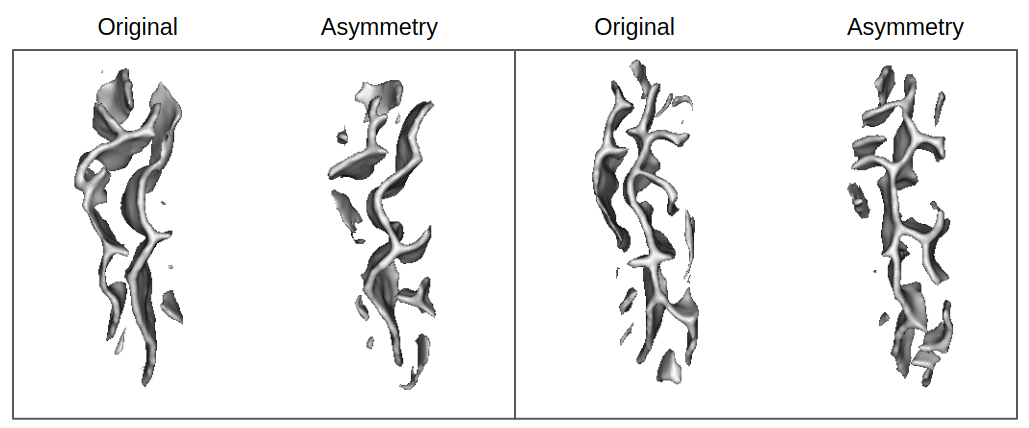}
    \caption{\textit{Asymmetry benchmark.} Visualization of the original sulcal pattern and its flipped contralateral version for two subjects.}
    \label{fig:description_asymmetry}
\end{figure}

Our second benchmark corresponds to the equivalent crop but in the left hemisphere. Left hemisphere distance maps are generated according to the same methodology as the right. Like our control crops of the right hemisphere, we computed a left central sulcus mask on the labeled dataset. To enforce the exact same crop size, we adapted the mask to match the adequate dimensions by adding or deleting a few voxels.
Once the crops were obtained, they were flipped. During training, the right central sulcus mask was applied on the fly. We emphasize that we did not use the interhemispheric plane-symmetric coordinates but a mask specifically designed for the left central sulcus. This is especially important since there is a slight asymmetry in the position of the central sulcus between the two hemispheres \citep{davatzikos_morphometric_2002}. An example is presented in Fig.\ref{fig:description_asymmetry}. 

\subsection{Identifying Outliers}
Once the model has learned a representation of the inter-individual variability, outliers identification can be performed at two levels. Traditionally, anomaly detection with AE is done based on the reconstruction error and an error map can be obtained comparing the input and the output \citep{schlegl_unsupervised_2017, schlegl_f-anogan_2019, pinaya_using_2018}. But one can also wonder about the distribution of outliers in the latent space. Are the outliers distributed differently? To answer this question, we investigated the detection power in the outliers' distribution in the latent space based on the reconstruction errors performed in the input space---which we call \textit{folding space} in our case, as we study folding patterns. For both approaches, control test images and outlier images (deletion benchmarks, asymmetry benchmark and interrupted sulci) are encoded and reconstructed by our trained model. 

\subsubsection{A specification on data}
As mentioned in \ref{training}, our control test set comprises 200 subjects. However, when studying our different outliers sets, data subsets were different since some subjects did not have any SS meeting the benchmark's criteria.
\begin{itemize}
    \item Deletion benchmarks: to avoid any bias, we used only control subjects with a simple surface meeting the benchmark's criteria for each benchmark. Therefore, for benchmark 200, we used 90 controls that have a simple surface between 200 and 500 voxels but that has not been erased, and 90 benchmark subjects, for whom simple surfaces were actually erased. Resulting in $n_{control}^{200}=n_{deletion}^{200}=90$, $n_{control}^{500}=n_{deletion}^{500}=34$, $n_{control}^{700}=n_{deletion}^{700}=54$ and $n_{control}^{1000}=75$ and $n_{deletion}^{1000}=76$.
    \item Asymmetry benchmark: all subjects have their asymmetric counterpart. Hence, 100 subjects were randomly picked among the subjects from the test set for whom we took their asymmetric version. Resulting in $n_{control}=n_{asymmetry}=100$.
    \item Interrupted central sulci: the whole test set is used as control data, leading to $n_{control}=200$ and $n_{interrupted}=7$.
\end{itemize}

\subsubsection{On the Latent Space}

\paragraph{A hint from the visualization}
For both of our benchmarks and the interrupted central sulci, we first sought to have a visualization of data distribution in the latent space. Therefore we projected encoded data into a smaller space of two dimensions with UMAP algorithm \citep{mcinnes_umap_2018, mcinnes_umap_2020}. This projection enables us to get a first hint as to how outliers are represented. 

\paragraph{Assessing the detection power on the benchmarks}
However, the UMAP algorithm drastically reduces dimensions, leading to some information loss. We tried to assess whether relevant information regarding folding patterns was encoded in the latent space. Therefore, we trained linear support-vector machines (SVM) \citep{pedregosa_scikit-learn_2011} on the latent codes with stratified cross-validation to classify between control data and benchmark. Performance is assessed based on the ROC curve.

\paragraph{Quantifying the marginality of interrupted central sulci}
As interrupted central sulci are very few, we cannot use classification as we did for the benchmarks. Classic machine learning out-of-distribution algorithms are more suited. Therefore, to quantify whether the interrupted sulci are likely to be detected from their location in this reduced space, we applied two classic algorithms, One-Class SVM (OCSVM) \citep{scholkopf_estimating_2001, pedregosa_scikit-learn_2011}, and isolation forest \citep{liu_isolation_2008, pedregosa_scikit-learn_2011} based on the data coordinates in the UMAP space. However, interrupted central sulci may not be the rarest pattern, and other folding configurations may be very scarce. Therefore, we also looked at control subjects repeatedly predicted as outliers by these algorithms.

\paragraph{Travelling through the latent space} 
Finally, to better understand the encoded properties and the learned representations, we leverage the generative power of the $\beta-VAE$. We computed average representations from different sets of data points, taking the mean for each dimension of the latent space. We then reconstructed these vectors. To further analyze the latent space, we traveled through it, going from one point, either the average pattern or a subject, to another point in the latent space, linearly interpolating vectors and reconstructing them.   

\subsubsection{On the Folding Space}
Outlier identification in the folding space relies on the model's error. The reconstruction errors' distributions were compared visually and assessed with the Kolmogorov-Smirnov test for the benchmarks and with the Mann-Whitney U-test for interrupted central sulci. For both cases, the null hypothesis was that the two distributions were identical. \\
The other strength of analyzing this space rather than the latent space is that the model's errors can help understand and locate the rare patterns' characteristics. To localize the errors, we commonly look at the residuals, which are the difference between the input and the reconstruction of the model. This corresponds to what the model has missed or added. To differentiate these two types of errors, we looked at them independently, computing the difference between the input and the output, i.e., the model's omissions, and between the output and the input, i.e., the model's additions. It is particularly interesting in the case of interrupted sulci, as we could expect that the model makes them continuous.

\subsection{Generalization to another region}
To assess the reproducibility in another region and to ensure that our framework is not limited to the central sulcus, we transposed our methodology to the isolated corpus callosum dysgenesis (CCD) which leads to a cortex anomaly located in the cingulate region. This disorder is a congenital malformation that results in a complete or partial absence of the corpus callosum. The corpus callosum is composed of fibers that connect the two hemispheres.

\subsubsection{Database}
The dataset includes 7 children between 9 and 13 years old presenting an isolated CCD and 7 matched control children \citep{benezit_organising_2015}. Among the patients, 3 present a complete agenesis, 3 a partial agenesis, and one a hypoplasia, corresponding to "a homogeneous reduction of the callosal size" \citep{tovar-moll_neuroplasticity_2007}. In this case, the corpus callosum is completely formed, but abnormally small \citep{bodensteiner_hypoplasia_1994}. For all children, the CCD was not associated with other malformations or developmental disorders. As presented before, we used T1-w MR images obtained from a Siemens Tim Trio 3T scanner with an isotropic resolution of 1mm. 

\subsubsection{Transposition of the method}
The described anatomical anomalies associated with CCD include "sulci radiating on hemisphere medial surface, complete or partial absence of the callosomarginal sulcus and of the cingulate gyrus" \citep{benezit_organising_2015}. Therefore, we transposed our method to the cingulate sulcus region. Using the same methodology as presented before, we computed a mask of the cingulate sulcus (gathering the calloso-marginal anterior and posterior fissure in the BrainVISA nomenclature), resulting in crops of dimensions 30 x 128 x 125 and 30 x 130 x 108, which were padded up to 32 x 128 x 128 and 32 x 136 x 112 respectively for the right and left hemispheres. We used the same data split as before. We used the hyperparameters obtained with the gridsearch on the central sulcus region for training. Choosing these parameters may lead to sub-optimal performances but enables us to have a first validation of our methodology. Analyses of the latent and the folding spaces are performed following the method described above for the central sulcus. Since the corpus callosum connects the two hemispheres, CCD can be studied equally in both hemispheres. Therefore, we conducted our experiments in the right and in the left hemisphere.

\section{Results}

\subsection{Training Results}
Each training lasted for approximately 1 hour on an Nvidia Quadro RTX5000 GPU. We obtained with our gridsearch $\beta$ = 2 and L = 75.

\subsection{Assessment on Synthetic Known Anomalies}

\subsubsection{On the Latent Space}

\begin{figure*}[ht]
    \centering
    \includegraphics[scale=0.28]{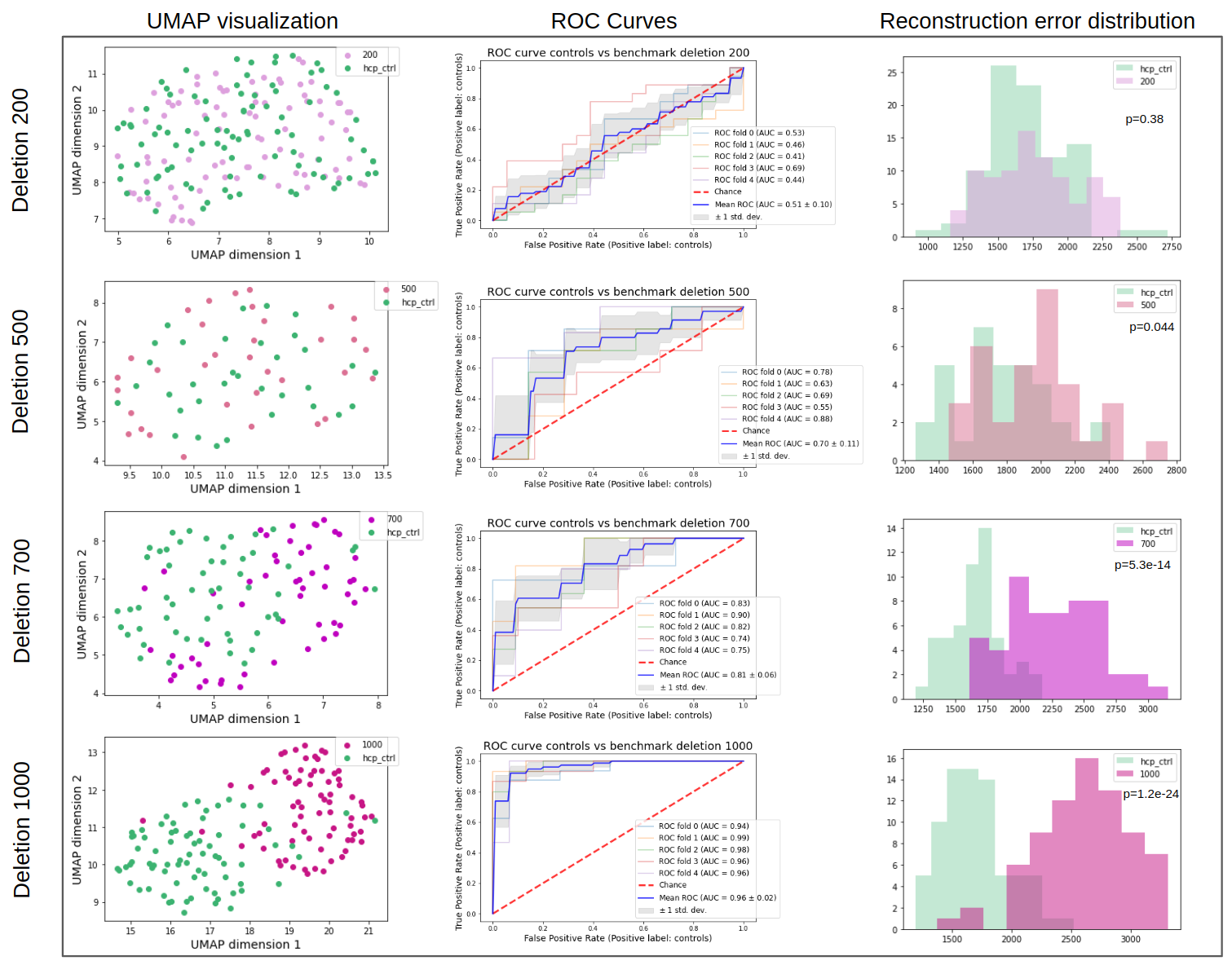}
    \caption{\textit{Deletion benchmarks results.} For each row, controls are represented in green and benchmark data in pink. Left column: UMAP projection of benchmark and control data. Middle column: ROC curves of classification of control and benchmark data. Right column: reconstruction error distributions and p-value of the Kolmogorov-Smirnov test with the null hypothesis that the two samples come from the same distribution.}
    \label{fig:results_deletion}
\end{figure*}

UMAP latent space visualizations for the four deletion benchmarks are presented in Fig.~\ref{fig:results_deletion}. For the benchmark 200, benchmark data are rather homogeneously distributed among control data, suggesting that simple surfaces of sizes between 200 and 500 voxels are too subtle to be encoded differently. Indeed, as shown in Fig.~\ref{fig:description_deletion}, small, simple surfaces can correspond to tiny branches that display a high variability in the population. Therefore these synthetic anomalies may be included in the normal variability. The distribution of benchmark 500 seems to be not completely similar to the control's, but the restricted number of subjects makes it hard to conclude. However, the trend becomes more pronounced for benchmarks 700 and 1000 where fake anomalies are gradually gathered and their distributions are different from the controls. These results are confirmed by the ROC curves (Fig.\ref{fig:results_deletion}). Even when using all the latent dimensions, classification results are very poor for benchmark 200 (AUC = 0.51), supporting that the deleted branches may be too melted into the inter-individual variability. Classification performances are also very low for benchmark 500 (AUC = 0.70). They start to be slightly better for benchmark 700 (AUC = 0.81) but are very good only for benchmark 1000 (AUC = 0.96).

\begin{figure*}[ht]
    \centering
    \includegraphics[scale=0.22]{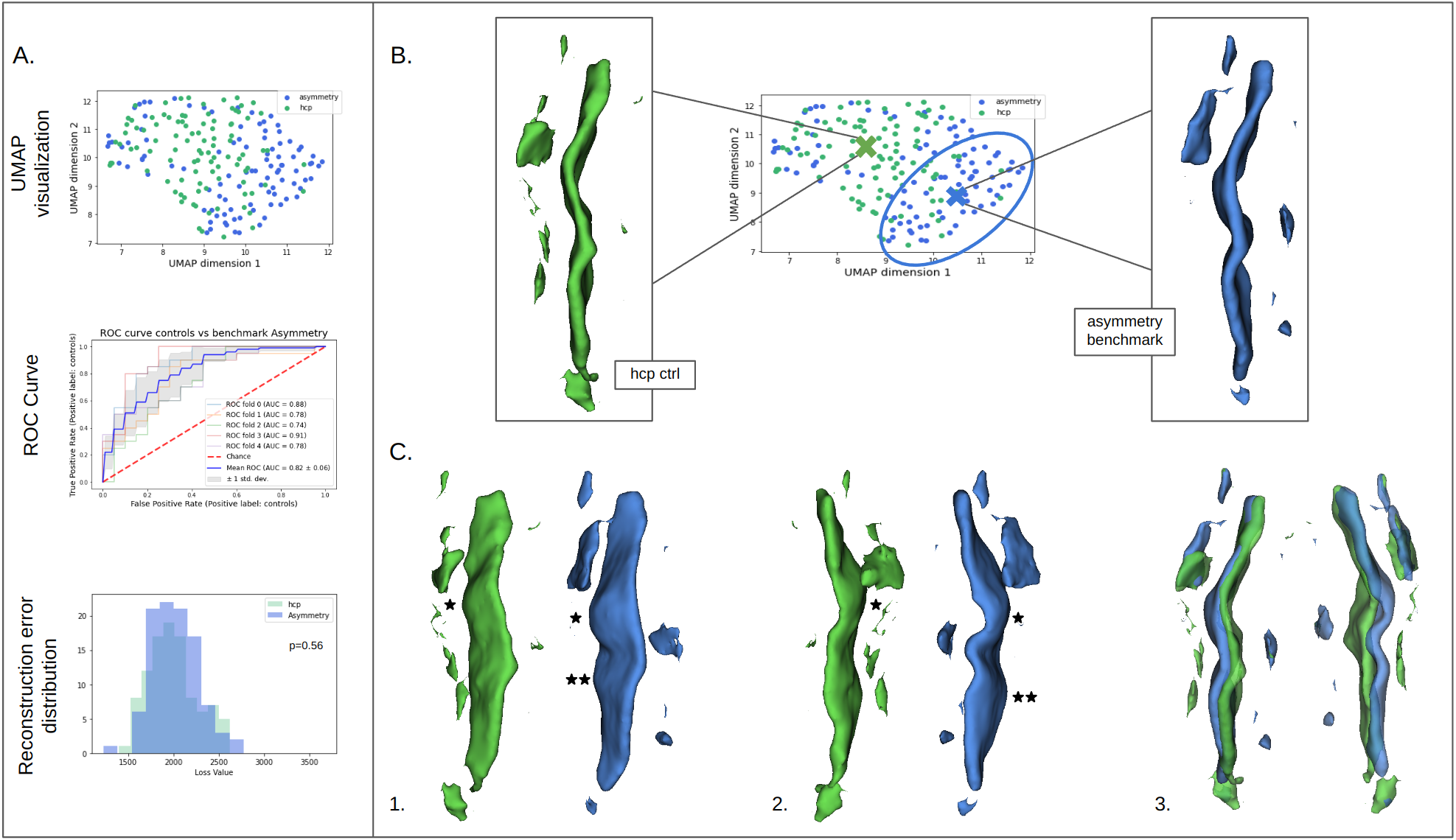}
    \caption{\textit{Asymmetry benchmark results.} Controls are represented in green and benchmark data in blue. A. UMAP projection of benchmark and control data, ROC curves of classification of control and benchmark data, and reconstruction error distributions. B. Averages for the control subjects, i.e. right hemispheres (in green), and for the highlighted asymmetry subjects, i.e. left hemispheres (in blue). These averages are also placed on the UMAP dimensions. C. 1. and 2. Respectively side and bottom views of the averages of B. The single star indicates a single-knob configuration, and the two stars indicate the second knob of a double-knob configuration. C. 3. Superposition of the two averages respectively in upper and bottom view.}
    \label{fig:results_asymmetry}
\end{figure*}

For the asymmetry benchmark, UMAP visualization demonstrates a good separation between the right and the left hemisphere (Fig.\ref{fig:results_asymmetry}A), which is verified by the classification of the whole latent space (AUC=0.82). These results suggest that specific shape features are encoded among other properties in the latent space.

\begin{figure*}[ht]
    \centering
    \includegraphics[scale=0.23]{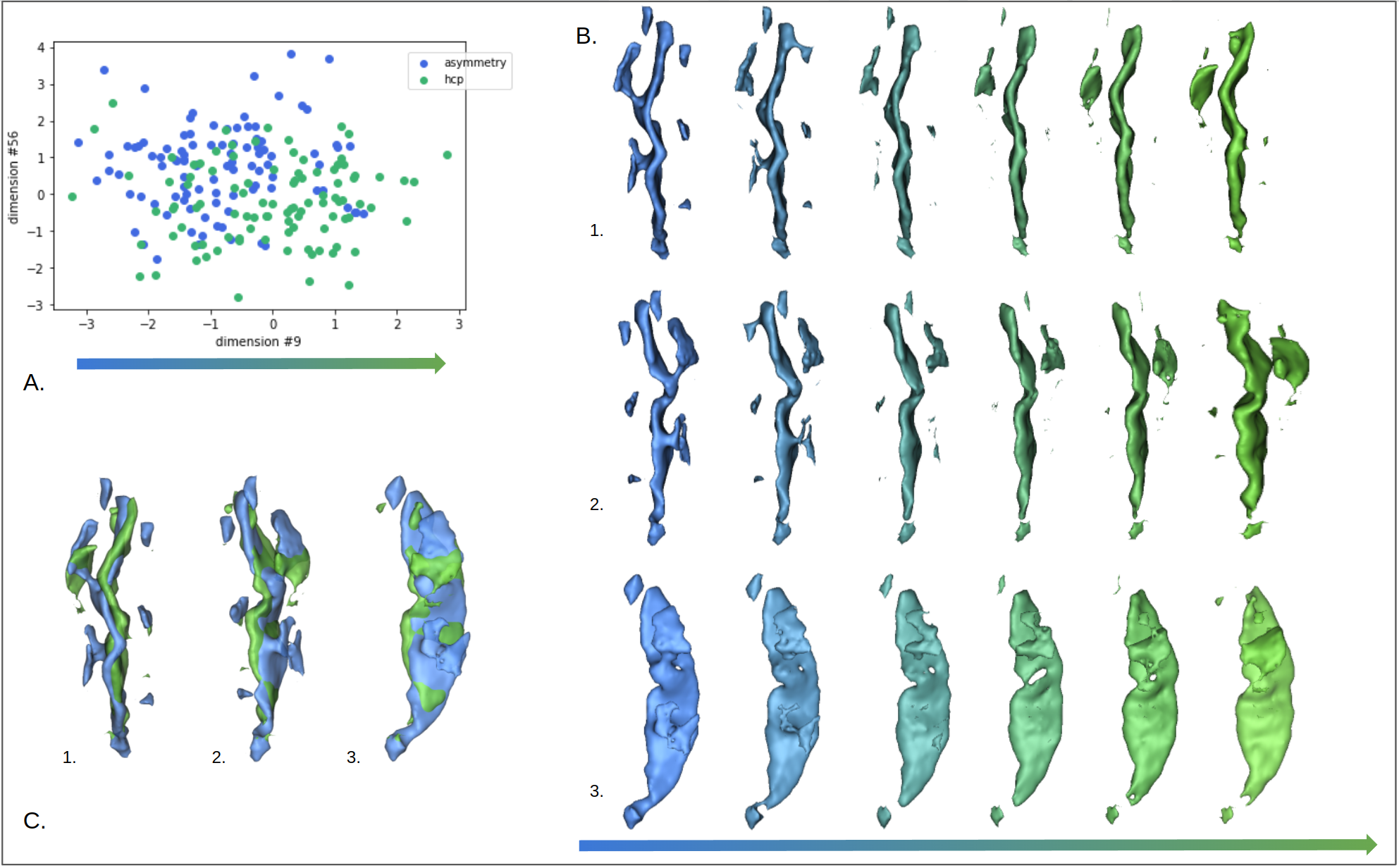}
    \caption{\textit{Travelling through the $9^{th}$ dimension of the latent space.} A. Visualization of controls and asymmetry benchmark according to the most important features of the classifier. B. Interpolations along the $9^{th}$ dimension. 1, 2, and 3, respectively correspond to the upper, bottom and side view of these interpolations. C. Superposition of extreme interpolations.}
    \label{fig:interpolation_asymmetry}
\end{figure*}

To better understand the asymmetry characteristics encoded by the model, we leveraged the generative power of our $\beta-VAE$. Fig.~\ref{fig:results_asymmetry}B. and C. show the average patterns for the right (green) and the left hemisphere (blue) as encoded by our model. The hand knob of the right central sulcus seems to be slightly higher and shallower than in the left hemisphere.
Moreover, the double-knob configuration appears more prominent in the left hemisphere. To further highlight the main differences between the two hemispheres, we selected the most important dimensions for the classifier, here dimensions 9 and 36. In Fig.~\ref{fig:interpolation_asymmetry}A., control and benchmark data are represented according to these two dimensions. Even if the separation is not well marked, we can observe a trend represented by the arrow. We tried to understand the features encoded by the 9th dimension. We took the average for all 75 dimensions of the latent space, and we traveled from the minimum to the maximum of the 9th dimension and reconstructed the resulting vector. Fig.~\ref{fig:interpolation_asymmetry}B. 1, 2, and 3 represent the reconstructions. These interpolations confirm the trend observed previously. We observe a double-knob configuration in the left hemisphere. The view from underneath and the side view enable visualizing the pli de passage frontal moyen (PPFM). A pli de passage is a gyrus that connects two gyri and which is buried in the depth of some furrows \citep{mangin_plis_2019}. Fig.\ref{fig:morphologist}B.1. and 2. propose a visualization of a "pli de passage" located in the central sulcus, the PPFM.  According to the different views from Fig.\ref{fig:interpolation_asymmetry}B. 1, 2 and 3, it seems that the PPFM is smaller in the right hemisphere and located higher in the central sulcus.

\subsubsection{On the Folding Space}
We then investigated whether the folding space, i.e, reconstruction errors, was relevant for identifying outliers. For deletion benchmarks, we observe a similar trend as in the latent space. For deletion 200, we cannot see a difference of distributions (p-value = 0.38). However, from deletion 500 we can see a stall with the deletion benchmarks having significantly higher reconstruction errors (p-values of 0.044, 5.3e-14 and 1.2e-24 for benchmarks 500, 700 and 1000 respectively) (Fig. \ref{fig:results_deletion}). On the contrary, for the asymmetry benchmark, there is no significant difference, nor a trend, in the reconstruction error distributions (Fig. \ref{fig:results_asymmetry}).

\subsection{Application on the Case of Interrupted Central Sulcus}

\subsubsection{On the Latent Space}
\begin{figure*}[ht]
    \centering
    \includegraphics[scale=0.21]{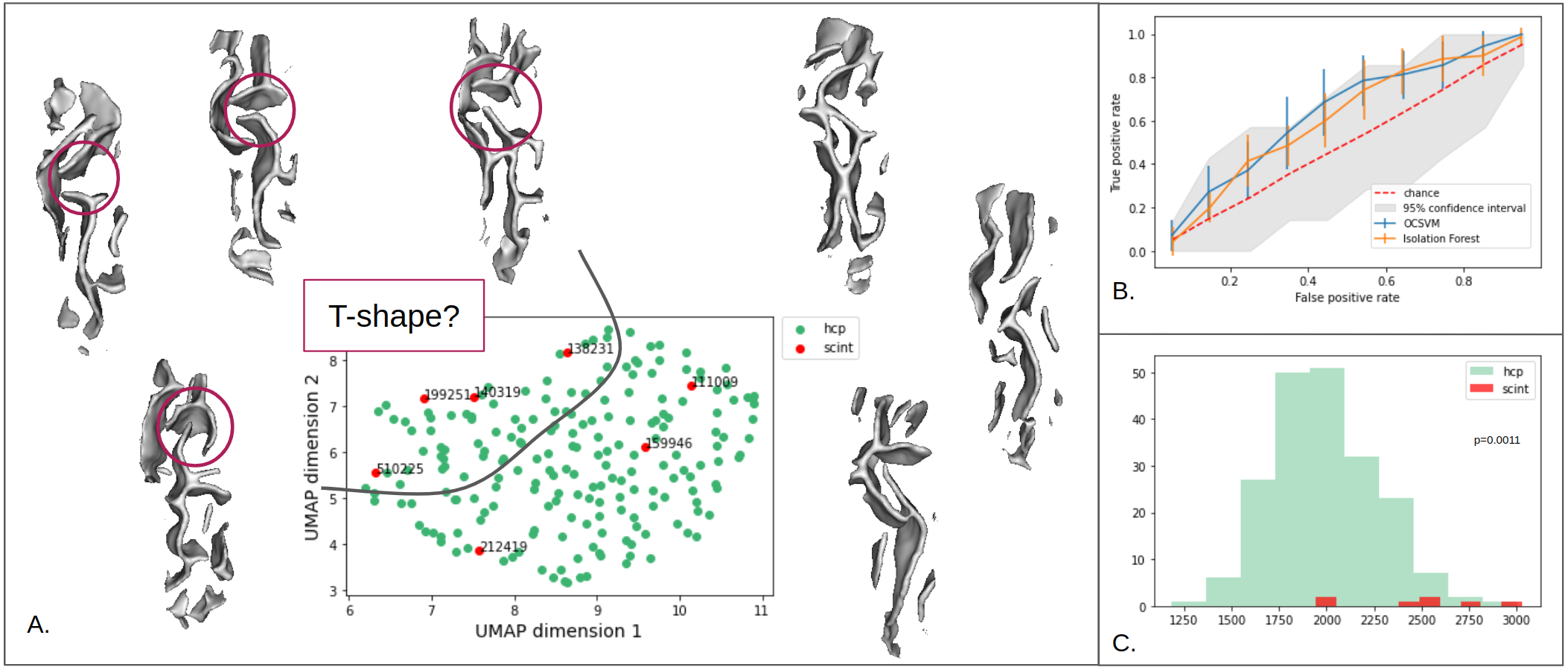}
    \caption{\textit{Interrupted central sulci on UMAP space.} A. Interrupted central sulci shape distribution in the UMAP space. The 3D folding patterns of the subjects are positioned according to their location in the UMAP space. For instance, the pattern located in the lower left corner corresponds to subject 510225 in the UMAP representation. Subjects with interrupted sulci on the upper left of the UMAP visualization seem to correspond to an interruption with a T-shape pattern. B. Outlier detection performances using OCSVM and isolation forest on the interrupted CS. C. Controls and interrupted CS reconstruction error distributions.}
    \label{fig:scint_umap}
\end{figure*}

The UMAP projection from the latent space is shown in Fig.~\ref{fig:scint_umap}A. On this distribution, we can observe that most interrupted central sulci are at the margin of the point cloud except for one. Thus, it appears that the representation learned by our model enables to project rare patterns at the margin of the population. Interestingly, when we look at the pattern of each one of the interrupted sulci, it seems that a specific pattern, the "T-shape" pattern \citep{mangin_plis_2019} is specifically located on one side of the representation.
Fig.~\ref{fig:scint_umap}B. shows the assessment of the marginality of the interrupted sulci based on an OCSVM and isolation forest. Error margins correspond to various UMAP projections, suggesting that the ability to detect interrupted CS in the UMAP space is very dependent on the UMAP projection. Interrupted CS detection is within the confidence interval, but the curves are close to the superior bound suggesting a tendency. However, interrupted CS positions in the UMAP space are not enough to detect them: detecting 5 interrupted CS out of 7 would lead to more than 40\% of false positives. Nevertheless, some other patterns considered as controls and detected as outliers might also be rare. 

\begin{figure*}[ht]
    \centering
    \includegraphics[scale=0.28]{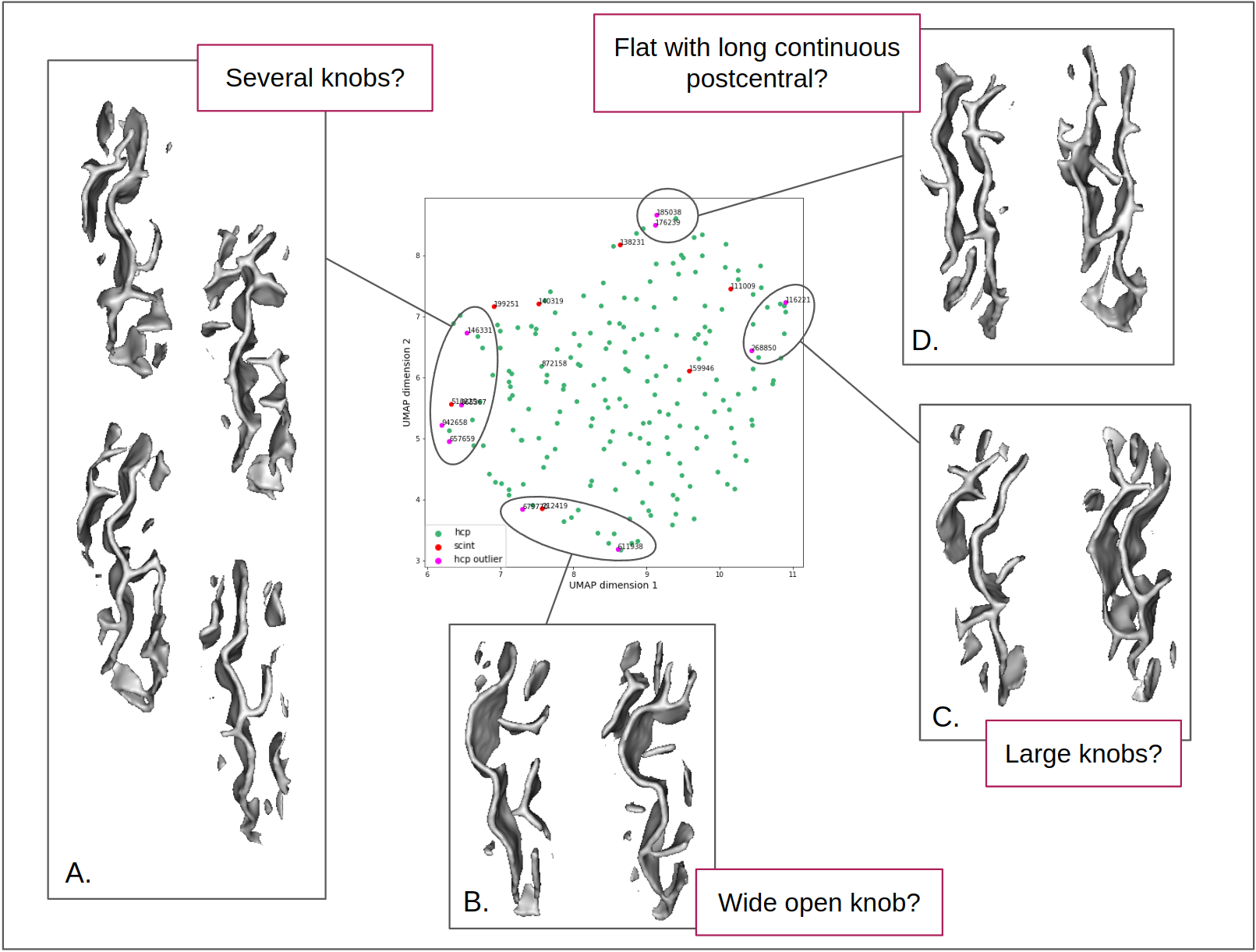}
    \caption{\textit{Control subjects identified as outliers.} A, B, C and D correspond to groups of visually similar patterns. The UMAP projection is the same as the one in Fig.\ref{fig:scint_umap}. Control subjects identified as outliers are in pink and subjects with interrupted central sulci are still represented in red.}
    \label{fig:ctrl_outliers}
\end{figure*}

\begin{figure*}
    \centering
    \includegraphics[scale=0.45]{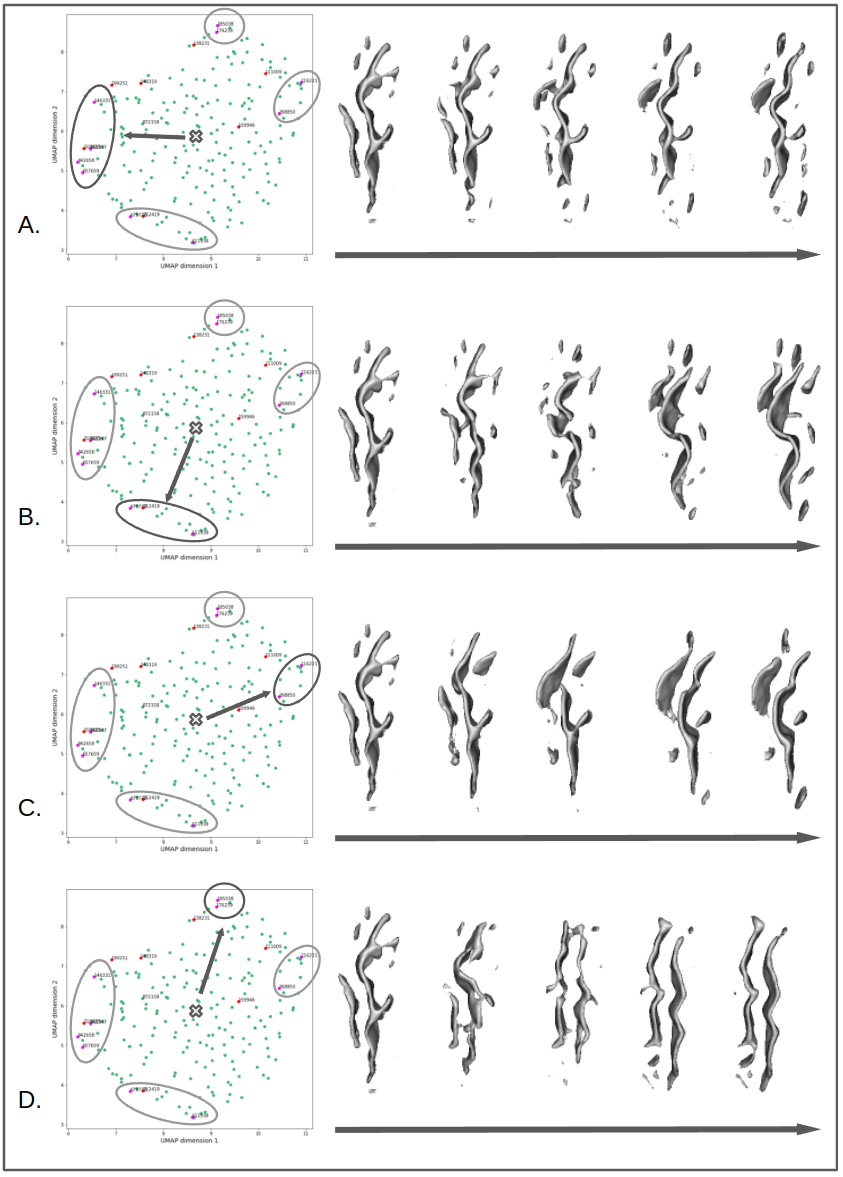}
    \caption{\textit{Travelling through the latent space from the centroid to the margin of the UMAP space.} The centroid is the centroid of HCP controls. Then, for each row, interpolations between the centroid and one of the patterns of each group are computed and then reconstructed.}
    \label{fig:ctrl_outliers_interpolations}
\end{figure*}

Fig.~\ref{fig:ctrl_outliers} presents the controls' patterns most often predicted as outliers by the OCSVM. First, we note that the outliers are logically located at the border of the distribution. Moreover, we observe distinct patterns in different regions of the UMAP space. We visually highlighted the subjects of the four regions. Analyzing the corresponding crops' meshes, we observe similarities within the groups. Group B seems to demonstrate a very wide open knob. In addition, the knobs are well defined by the upper and the bottom part of the sulcus. On the contrary, the sulci of group C appear to have larger knobs than usual but they show more continuity with the upper and the bottom parts. The pattern of group D seems to correspond to a rather flat central sulcus with a close, long and continuous postcentral sulcus. The shape characteristics of A are less obvious but the sulci give the impression of having several small knobs, two or even three in the two bottom cases and a small part of the precentral inferior opposite to an upper part of the postcentral sulci. Fig.~\ref{fig:ctrl_outliers_interpolations} provides a better understanding of these features. For each pattern, we go from the centroid to one of the subjects in each group by interpolating and generating samples. Fig.~\ref{fig:ctrl_outliers_interpolations}A. presents the interpolations from the centroid to the several-knobs pattern. We gradually see the upper part of the hand knob curving and becoming more pronounced until forming a first knob at the top of the sulcus. Another knob in the bottom part appears similarly. Likewise, patterns B, C and D vary progressively until they match the centroid's shape.

\subsubsection{On the Folding Space}
When analyzing the detection power on interrupted CS in the folding space, we first note that the reconstruction errors' distributions seem to be different between HCP controls and interrupted CS (p-value = 0.0011). This result suggests that our model has more difficulties to reconstruct the input and that reconstruction error could constitute a relevant metric to detect rare or abnormal patterns. However, having only seven subjects strongly limits our conclusions and this should be replicated with more data.  

\begin{figure*}[h]
    \centering
    \includegraphics[scale=0.18]{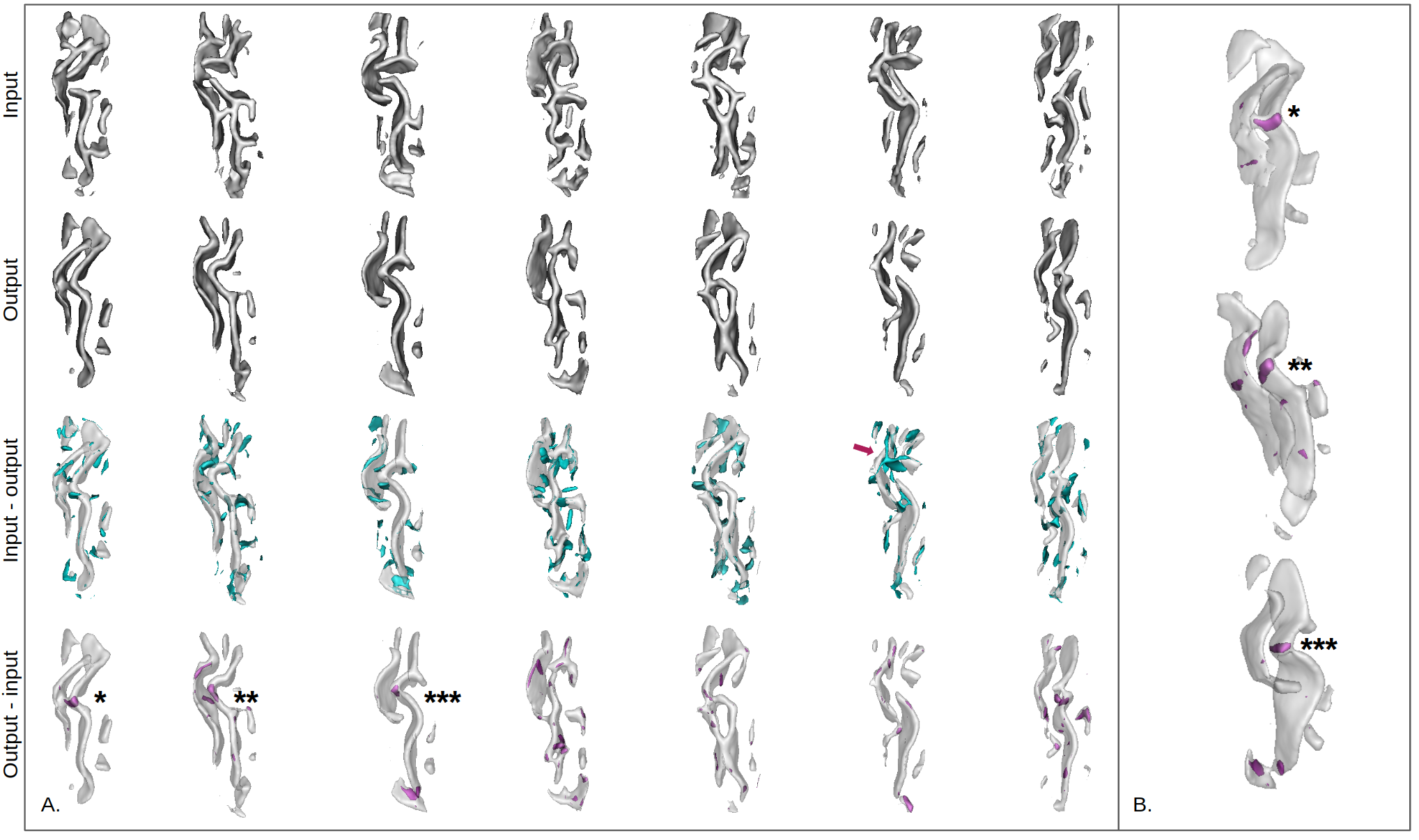}
    \caption{\textit{Reconstructions and residuals for all seven interrupted sulci.} A. For all rows, distance maps are converted to meshes for an easier visualization. First row: input data. Second row: reconstruction of the model. Third row: Reconstruction of the model with the difference between the input and the output, i.e. the model's omissions (in blue). The purple arrow highlights an omission corresponding to a perpendicular branch pointing toward the frontal cortex. Last row: Reconstruction of the model with the difference between the output and the input, i.e. the model's additions (in purple). B. Rotated view of the reconstructions represented with asterisks in the last row of A.}
    \label{fig:scint_reconstruction}
\end{figure*}

Observing the reconstructions and the residual maps of Fig.~\ref{fig:scint_reconstruction} gives clues into the way our model has encoded the interrupted CS. First, we can note that the reconstruction quality is quite good visually. The model's omissions appear to be quite noisy (blue small fold pieces). The arrow points out an omission beyond the noise which corresponds to a perpendicular branch pointing toward the frontal cortex. Such a pattern might be an uncharacteristic configuration. It is interesting to note that in six out of seven cases, the model transformed interrupted sulci into continuous patterns. This is highlighted by the "output-input" visualizations. Unlike the omissions, the model additions are rather localized. Moreover, the asterisks show where the model has filled the interrupted sulci. Such visualization could be useful to identify rare patterns like interruptions or perpendicular branches.  

\subsection{Application to corpus callosum dysgenesis}

\subsubsection{On the Latent Space}
\begin{figure*}[h]
    \centering
    \includegraphics[scale=0.22]{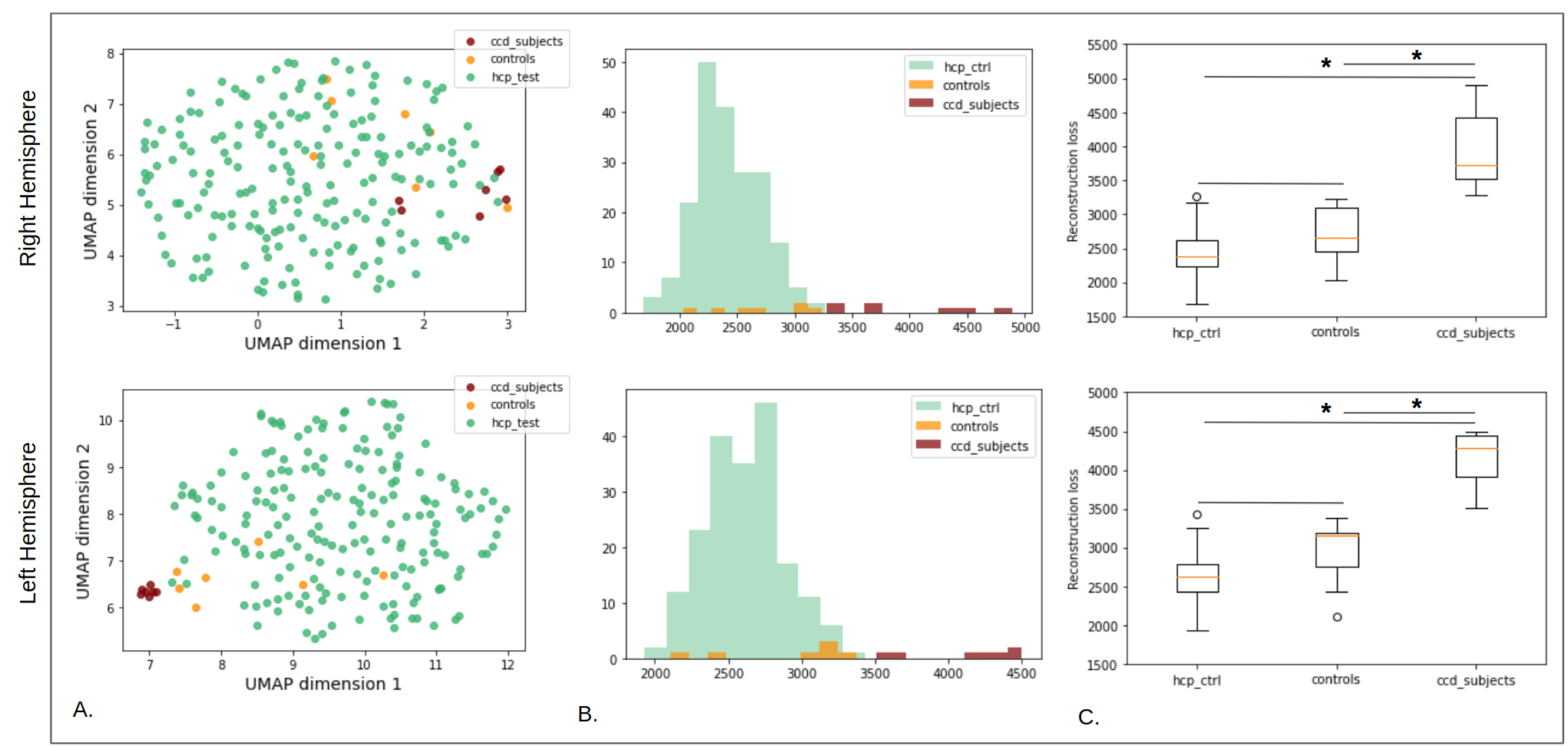}
    \caption{\textit{Results on corpus callosum dysgenesis (CCD) subjects.} First row: right hemisphere. Bottom row: left hemisphere. For both rows:  A. UMAP projections of CCD subjects, control children and HCP test. B. Reconstruction error distributions for the CCD subjects, control children and HCP test. C. Reconstruction error variations for the CCD subjects, control children and HCP test. Significant differences between populations according to the Mann-Whitney test are indicated with an asterisk.}
    \label{fig:results_ccd}
\end{figure*}

We first compare distributions of CCD children (n=7) with control children (n=7) acquired in the same conditions and with HCP adult subjects (n=200). UMAP projections, presented in Fig.\ref{fig:results_ccd}A., give different results depending on the hemisphere. For the right hemisphere, it seems that most children controls are included in the distribution of adult controls (hcp\_test in green). Five out of the seven subjects having a CCD are located at the margin of the controls, suggesting that their latent representation differs from the average cingulate sulcus pattern. However, two subjects, one with a complete and one with a partial agenesis, are in the middle of the controls.
In the left hemisphere, only three control children are clearly in the control adult distribution. The other four are closer to the CCD subjects but they seem to be still distinct. Indeed, CCD subjects are gathered very close to each other. This could be due to the fact that there may be an age effect between children's and adults' brains or a site effect (different scanners, resolution), which we tried to reduce by using skeleton-based images but which may still remain. Nevertheless, we can still observe a difference in distribution between control children and CCD subjects. 

\subsubsection{On the Folding Space}
Regarding reconstruction error distributions (Fig.\ref{fig:results_ccd}B.), we observe for both hemispheres that control children seem to have the same distribution as adult controls, which is confirmed by Fig.\ref{fig:results_ccd}C. (p-value=0.034 and 0.017 respectively for right and left hemisphere). On the contrary, CCD subjects present higher reconstruction errors that are significantly different from both HCP controls (p-value=3.6e-06 for the two hemispheres) and children controls (p-value=0.0011 for the two hemispheres). Therefore, it seems that there is a complete individual separability of the CCD patients which is very promising and should be replicated with more data.

\begin{figure*}[h]
    \centering
    \includegraphics[scale=0.15]{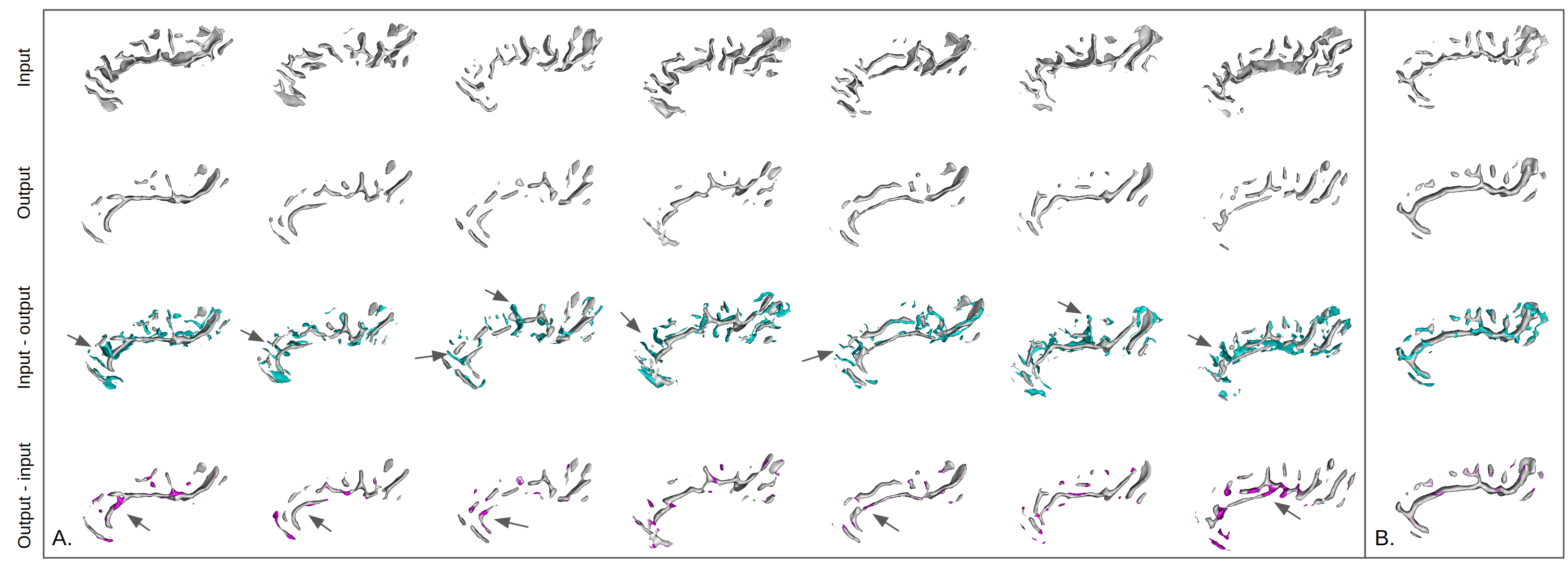}
    \caption{\textit{Right cingulate sulcus reconstructions and residuals for the CCD subjects and one control.} A. CCD subjects. B. One control subject from the same cohort. For both A. and B.: each column corresponds to a subject. For all rows, distance maps are converted to meshes for an easier visualization. First row: input data. Second row: reconstruction of the model. Third row: Reconstructions of the model with the difference between the input and the output, i.e. the model's omissions. Last row: Reconstructions of the model with the difference between the output and the input, i.e. the model's additions. The arrows highlight interesting features added or missed by the model.}
    \label{fig:reconstructions_ccd}
\end{figure*}

The reconstructions presented in Fig.\ref{fig:reconstructions_ccd} highlight the singularities of CCD. The model's additions mostly make the cingulate more continuous than initially. The model's omissions are mainly small branches perpendicular to the cingulate sulcus that are radially oriented.   

\section{Discussion}

This work proposed a methodology to study rare folding patterns which was applied to the central sulcus region and to a described rare pattern, interrupted central sulci. Specifically, we represented folding patterns with distance maps and leveraged the generative power of the $\beta-VAE$ to have a better understanding of the learned representations. In addition, we proposed a way to study the granularity of deviations that can be identified and we brought to light several rare patterns in the region. We also compared the identification power of both the latent space and the folding space. Finally, we assessed the generalization of our methodology on a developmental anomaly in another region.

\subsection{Latent Space and Folding Space, Two Complementary Information}
In many anomaly detection works applied to medical images, the detection is performed based on the reconstruction error rather than in the latent space \citep{schlegl_f-anogan_2019, baur_autoencoders_2020, behrendt_unsupervised_2022, tschuchnig_anomaly_2022}. However, both of these spaces have their interest and could bring complementary information. In our work, we studied four types of rare patterns, two synthetic types, deletion and asymmetry benchmarks, and two actual rare patterns. These four categories differ from control data by their own characteristics and thus help to study the granularity detected, that is to say, the typology of rare features that can be identified. For instance, the asymmetry benchmark includes more double-knob configurations. Depending on the size of the deleted simple surface, deletion benchmarks represent different features: benchmarks 200 and 500 represent mainly a missing branch with increasing size, which may represent the normal variability of branches. Benchmark 700 could look like an interrupted sulcus in some cases or in others, like benchmark 1000, an unlikely configuration. Interrupted central sulci present a clear interruption and a rare arrangement of the shapes forming the central sulcus. Last, CCD subjects demonstrate a missing sulcus or missing sulcal parts and branches with different orientations. 

These different kinds of deviations from the norm provide clues to the characteristics of rare patterns that can be identified respectively in the latent space or in the folding space of our model.
As a matter of fact, the identification performances in the latent and in the folding space vary depending on the kind of patterns. For deletion benchmarks, the folding space, based on the reconstruction error, seems to enable the identification of unusual patterns from smaller modifications: different distributions are observed from 500 deleted voxels. Whereas in the latent space, the detection requires at least 1000 deleted voxels. Likewise, for the interrupted central sulci, despite the small number of samples, their detection seems to be easier on the basis of reconstruction error than in the latent space. Similar results were obtained on CCD subjects even if the latent representation was encouraging. In return, the error distribution of the asymmetry benchmark is not different from that of the controls, but the benchmark is well detected in the latent space. Therefore, the latent space could be more sensitive to shape arrangements than the folding space. The lack of difference in the error distributions may be due to the fact that the voxel-to-voxel differences between the right and left central sulci are local and subtle and could be embedded in the normal variability. In addition, the reconstruction error is for the entire image. Therefore, in the case of small and very local deviations from the norm, the reconstruction error alone is likely to be insufficient. A way to limit such effects could be to use a more local error, applied to sub-regions or patches for instance. 

The difference in the outlier detection performance may also lie in the way our model encodes the outliers. Based on our results, we can consider several cases. First, a rare configuration is represented by several samples present in the training set. This would be the case with the asymmetry benchmark. Indeed, there are more double-knob configurations in the left hemisphere but single and double-knob patterns coexist on both sides. In such a case, the distribution support of the left and right hemispheres are the same, but the densities differ, which could lead to a projection of the outlier at the margin of the latent space but to a good reconstruction. Second, the rare configuration is almost never represented in the training set and the model has not detected and thus encoded its local specificity. Then, the subject would be encoded with a "default" representation and projected in the middle of the other subjects. This would be consistent with the results of \citep{guillon_detection_2021}, where major anomalies (different parts of the brain from the one considered in the train set) were projected in the middle of the point cloud and reconstructed as the average reconstruction. It could be the case of the benchmark deletion 500 and of the interrupted central sulcus that is projected in the point cloud. Last, the outlier configuration is almost never represented in the training set but the model has detected the rare characteristic. The subject is then projected at the margin of the point cloud and the decoder has not learned this part of the latent space leading to a poor reconstruction (interrupted central sulci, CCD subjects).

Nevertheless, in all cases, a strength of the folding space is the possibility to localize the reconstruction errors and, in some cases, the unusual features. If not too noisy, reconstruction errors can be very informative. 
%In our work, as we are using distance maps, to observe the folding in 3D we keep only the voxels near the folds and for the errors we keep only the largest differences between the distance maps. A side consequence 
%using distance maps that we visualize in 3D enables to threshold the errors, getting rid of small errors. 
For example, in the case of interrupted central sulci, looking at the model's addition permits clearly localizing what is atypical in a subject (Fig.\ref{fig:scint_reconstruction}). Similarly, in the case of the CCD subjects, the reconstruction errors highlight the presence of radial small branches that are typical of this brain disorder \citep{benezit_organising_2015}. 
But some noise remains, and it might be interesting to add an additional constraint to represent only errors that correspond to a minimum number of contiguous voxels. This could lead to a good explanation of the abnormality which is of major importance in the field and especially when applied to medical images. Other explanation methods exist, directly on the network such as Grad-CAM \citep{selvaraju_grad-cam_2020} or on an OC-SVM applied on the learned features \citep{sohn_learning_2022} for instance; but the use of the reconstruction error is immediate and easy to implement. 
Hence, the latent space and the folding space, based on reconstruction error, can provide complementary information and both can be used to identify rare patterns.

\subsection{Data size limitations and unknown number of rare patterns}
The method should be further qualified because of the low number of our examples of rare patterns. While the study of a known rare pattern is interesting and important, having only seven samples severely limits our conclusions. Similarly, the poor results of benchmarks 500 and 700 in the latent space could be due to their small size, and having larger benchmark datasets could lead to increased performances.

%On the other hand, we mentioned in the introduction the existence of both global and local abnormal folding patterns. In this work, we chose to study regional patterns
Also, we assessed our method in the CS area on the benchmarks and on an existing rare pattern, but because few rare patterns have been described in this region, there may be other rare configurations in what we consider the control population. For instance, three morphologic variants in the central sulcus region have been introduced, representing 2.9\%, 7.0\% and 1.8\% of the studied population, opposed to 78.2\% of "omega" shape, i.e. the central sulcus knob and 10.1\% of "epsilon" shape which corresponds to the double-knob configuration \citep{caulo_new_2007}. This multiplication of rare patterns in the populations would make the identification of interrupted central sulci more difficult.

\subsection{Relevance of synthetic benchmarks}
Moreover, we can wonder about the relevance of our synthetic benchmarks. Although synthetic rare patterns are of high interest as they enable to quantify the performances on different degrees of deviations from the norm, few works have been interested in them to our knowledge \citep{guillon_detection_2021, meissen_pitfalls_2022}. But the use of fake deviations raises the question: do they constitute adequate rare patterns? Few studies introduced rare folding patterns based on the arrangement of their shapes such as the PBS \citep{mellerio_power_2014}, an interrupted central sulcus \citep{mangin_plis_2019} or a flat central sulcus \citep{sun_congenital_2017}. Here, we emphasize their advantage in the study of our understanding of the brain: they are evidence of neurodevelopmental processes and then stable throughout life. But other abnormal sulcal features have been studied and found to be important and correlated with neurodevelopmental disorders, such as the depth, which demonstrated anomalies in autism spectrum disorder \citep{nordahl_cortical_2007,dierker_analysis_2015} or Williams syndrome \citep{essen_symmetry_2006} for instance. Despite being another subject of study, a benchmark corresponding to central sulcus depth variations could be interesting to assess whether our framework can be extended to detect such anomalies.

Regarding the current benchmarks we use, we said that small erased SS could remain undetected as this deletion could be embedded in the normal variability. However, there may be several categories of deletion deviations. Some may be minor, as a small SS representing a tiny branch. On the opposite, some small SS, for instance one corresponding to depth change, representing the presence of a pli de passage, and thus leading to an interrupted central sulcus would be expected to be a major feature of the topology. Hence, our criterion, only based on the size of the SS may be insufficient and it could be interesting to add another one, such as topological criteria.

In any case, having an unusual feature (e.g., a missing simple surface or unusual depth) that can be incrementally increased, or comparing several types of features, helps characterize the detection power of a model and the features likely to be detected.

\subsection{Learning Relevant Representations}
When dealing with sulcal patterns and their high complexity, it may be easier to use representations of the folding which attempt to gather several subjects with similar patterns. Local averages of sulci, also called moving averages, enable to concentrate on the main features of the different patterns and are thus very useful to analyze folding patterns \citep{sun_effect_2012, de_vareilles_shape_2022, foubet_comparison_2022, guillon_unsupervised_2022}. From a graph-based representation of the sulci, the identification of patterns can be done after computing similarity and applying a clustering \citep{meng_discovering_2018}. Our approach proposes another method to learn sulcal representations. From our cropped distance maps, the $\beta-VAE$ learns a mapping to a latent representation which can then be reconstructed. Therefore, rather than explicitly computing pairwise similarity between the subjects, gathering them, and then analyzing the patterns, we hope that our $\beta-VAE$ directly learns shapes that can be combined and arranged in patterns. The representations learned by our model seem to be relevant and consistent with some morphological characteristics of the central sulcus area. \\
First, the reconstruction of the average representation of the right central sulcus is composed of an upper knob whereas the left average tends more towards a double-knob configuration (respectively green and blue sulci in Fig.\ref{fig:results_asymmetry}B). This is one of the main known asymmetries in terms of patterns and it appears early in the development. It has been detected in infants of 30 weeks postmenstrual age \citep{de_vareilles_shape_2022} and in adults \citep{sun_effect_2012}. \\
We also observed differences in terms of curvature of the hand-knob, with a hand-knob more pronounced in the left hemisphere than in the right (Fig.\ref{fig:results_asymmetry}B.). Considering that we study a right-handed population, this could be related to handedness. With the lateralization of the hand motricity, we expect the motor area of the right hand in the left hemisphere and particularly the precentral gyrus to be more developed for right-handed subjects, pushing backward the upper part of the central sulcus which would result in a knob more pronounced. This interpretation is consistent with a study on one-handed subjects that showed that subjects born without a hand had a flatter central sulcus contralateral to the missing hand \citep{sun_congenital_2017}.

Another interesting property that was successfully encoded is the PPFM. This pli de passage was first described in 1888 \citep{broca_memoires_1888} and has been a source of growing interest due to its link with the motor hand area \citep{boling_localization_2004} and in the context of understanding the formation of the knob regarding evolutionary questions \citep{hopkins_evolution_2014}. Our model was able to encode the PPFM in the latent space as well as its asymmetry characteristics. Indeed, we observed that the PPFM is smaller in the right hemisphere which corresponds to central sulcus depth variations described in \citep{amunts_asymmetry_1996}. This is also consistent as the PPFM has been correlated to the hand. Therefore, right-handed subjects tend to have a more developed hand area in the left hemisphere and thus a larger PPFM. \\
Hence, our latent space has learned relevant \textit{normal} characteristics that are consistent with the region's morphology. It has also enabled to propose four other groups of likely rare patterns (Fig.\ref{fig:ctrl_outliers}). The pattern representing a rather flat central sulcus is indeed a non-typical configuration. Less than 2\% of the studied subjects were reported to have such a configuration in \citep{caulo_new_2007}. Moreover, flat central sulci appeared as the most important feature when comparing controls to congenital one-handed subjects who tended to demonstrate flatter central sulci \citep{sun_congenital_2017}, confirming that flat central sulci are less frequent patterns. The groups representing large knobs and wide open knobs (Fig.\ref{fig:ctrl_outliers}B. and C.) are also an atypical configuration that is present at one extremity of the axis representing the most extreme variations in Human and is closer to configurations we observe in Chimpanzees \citep{foubet_comparison_2022}. 

\subsection{Generative power of $\beta-VAE$ and comparison with other strategies}

Since our proposed framework is able to encode relevant features regarding folding patterns, the generative power of the $\beta-VAE$ can be exploited. Indeed, reconstructions and interpolations are tools to understand the folding variability. We have just mentioned that the learned patterns were relevant and consistent with those obtained by other methods, but our method has the advantage of being able to reconstruct and interpolate. For instance, interpolations along the main axis of asymmetry variations highlight the evolution from a right to a left hemisphere.
It can also be useful to understand the folding process and in particular the formation of interrupted central sulci. 
As a matter of fact, on Fig.\ref{fig:ctrl_outliers_interpolations}C., an interruption of the central sulcus happens when interpolating from the central subject to one control outlier. When observing the PPFM, we can see that the PPFM increases until reaching the surface of the brain and thus interrupting the central sulcus. Jointly, the inferior sulcal part connects to the precentral sulcus. Such observations may provide additional clues in our understanding of the folding processes.

But other deep learning models could be interesting to study folding patterns. For instance, $\beta-VAE$ reconstructions are known to be blurry contrary to GAN's. Currently, this shortcoming is limited as we seek to have a simpler representation of folding patterns, still, for more subtle details, another model may be better suited. In addition, in the anomaly detection field, models that add constraints on controls distribution are quite appealing. For instance, deep One-Class Classification and its derivatives have been proposed to push control data into the smallest hypersphere in the latent space \citep{ruff_deep_2018}. This could help increase the detection performance in the latent space. Nevertheless, no matter the architecture or the framework, an important limit to understanding what our model has really encoded is the high number of latent dimensions.

One can also wonder about the representation of the folding patterns. In the introduction, we mentioned two main strategies: clustering and manifold. Usually, these two approaches are applied to a continuous space. Nevertheless, if we consider sulcal shapes as symbolic entities that can be combined and arranged, we could represent folding patterns based on a discrete space rather than a continuous one. As such, VQ-VAE \citep{van_den_oord_neural_2017} seems to be an interesting representation to compare with our present results.

Finally, this framework of outlier detection based on training on control subjects alone may be sensitive to outliers present in the training set. Having a contaminated dataset could severely limit the detection performances, at least in the folding space which is based on the reconstruction error. It has been reported in a brain tumor detection problem that having 3\% of outliers in the training set (about 1000 samples) leads to a decrease of 5\% of the AUROC and to a 13\% decrease if the contamination reaches 12\% of the training set \citep{behrendt_unsupervised_2022}. Therefore, one serious shortcoming of our paradigm is that we do not know the outliers we are looking for. Applying our framework to a control population alone in order to bring out rare patterns may limit the different patterns that can be identified. A way to tackle this issue and to increase the patterns detected would be to exploit the presence of outliers in the training set as proposed in \citep{qiu_latent_2022}. In their technique, the authors introduce an iterative joint training where they assign labels (anomalous or control) to the examples, and then optimize the network's parameters to better identify the anomalies.  Such a method could also enable to project the outliers more at the margin of the latent space. The impact of the presence of outliers during training on the latent space has not yet been investigated to our knowledge. If, as we suggested before, outliers present in the training phase are encoded at the margin of the distribution, i.e. in a different area of the latent space, it could be interesting to deepen our analysis, based on clustering for instance.

\subsection{Generalization of the approach: towards an analysis of the whole brain?}

This work has shown that our approach had successfully encoded some relevant features of the folding patterns in the central sulcus region but it is attractive to think about the behavior and results we could obtain in other parts of the brain. Here, we assessed the generalizability of the framework on another dataset and another region. Our results suggest that our method can well transpose in other brain regions. Specifically, even if we use the hyperparameters ($\beta$ and L) optimized for another area, the learned representations still enable us to distinguish between control and outlier subjects. This is all the more interesting that the two studied regions are rather different. The central sulcus is one of the first folds to form and is rather stable, contrary to the cingulate region that is more variable \citep{sun_constructing_2009}. Therefore, it seems that no matter the folding variability of the zone, our framework can be applied.
This encouraging result raises a question regarding the procedure to adopt to extend our analysis to the whole brain. A way could be to define a set of regions, consistent with the cytoarchitecture and function and to train our $\beta-VAE$ on each region. In particular, some areas seem to be interesting from a clinical point of view \citep{provost_paracingulate_2003, yucel_morphology_2003,gervais_abnormal_2004, borst_folding_2014, hotier_social_2017}. Our future works may thus focus on proposing an adequate methodology to tackle the whole brain. 

On the other hand, when we applied this framework to CCD subjects, we also operated a domain shift. Indeed, the dataset to explore included exclusively children while the $\beta-VAE$ was trained on young adults. Despite folding patterns being reported as trait features \citep{cachia_longitudinal_2016}, such an age variation may have an impact. In addition, beyond dealing with children, the site and the scanner are different. Such differences have been reported to affect the generalizability and the performances on various targeted tasks. In terms of distributions in the latent space, despite the fact that the distribution of the controls does not seem to completely overlap the distribution of the HCP controls, the patients still seem to present a different distribution than both controls' populations. Moreover, the domain shift does not seem to have an effect on the folding space where controls reconstruction errors are not significantly different from HCP contrary to CCD subjects that have significantly higher reconstruction errors. 
However, having only seven subjects makes it difficult to conclude on the importance of these age and site effects for our task. We will explore these questions in further studies. \\

To conclude, this study proposed a framework to identify rare and abnormal folding patterns based on the modeling of the inter-individual variability. With a new representation of folding patterns, we proposed a model that was able to encode relevant folding characteristics. The use of synthetic rare patterns enlightened the identification power of our model on both the latent space and the folding space. Finally, we successfully generalized our approach to another clinical brain anomaly in the cingulate region. Our results open up several avenues of work such as the definition of new synthetic benchmarks that match the characteristics of other known anomalies, the use of other deep learning models that exploit the presence of outliers in the training set, or the use of our framework to better understand the folding process.

\section{Funding}
This work was supported by the European Union’s Horizon 2020 Research and Innovation Programme under Grant Agreement No. 945539 (HBP SGA3), the ANR-19-CE45-0022-01 IFOPASUBA, the ANR-14-CE30-0014-02 APEX the ANR-20-CHIA-0027-01 FOLDDICO.

\section{Acknowledgments}
The authors thank G. Dehaene for her involvment in the scanning procedure of the children cohort and C. Langlet for his help in the discussions about this study. Data were provided in part by the Human Connectome Project funded by the NIH.

\bibliography{rare_folding_identification}

\begin{thebibliography}{70}
\expandafter\ifx\csname natexlab\endcsname\relax\def\natexlab#1{#1}\fi
\providecommand{\url}[1]{\texttt{#1}}
\providecommand{\href}[2]{#2}
\providecommand{\path}[1]{#1}
\providecommand{\DOIprefix}{doi:}
\providecommand{\ArXivprefix}{arXiv:}
\providecommand{\URLprefix}{URL: }
\providecommand{\Pubmedprefix}{pmid:}
\providecommand{\doi}[1]{\href{http://dx.doi.org/#1}{\path{#1}}}
\providecommand{\Pubmed}[1]{\href{pmid:#1}{\path{#1}}}
\providecommand{\bibinfo}[2]{#2}
\ifx\xfnm\relax \def\xfnm[#1]{\unskip,\space#1}\fi
%Type = Article
\bibitem[{Amunts et~al.(1996)Amunts, Schlaug, Schleicher, Steinmetz,
  Dabringhaus, Roland and Zilles}]{amunts_asymmetry_1996}
\bibinfo{author}{Amunts, K.}, \bibinfo{author}{Schlaug, G.},
  \bibinfo{author}{Schleicher, A.}, \bibinfo{author}{Steinmetz, H.},
  \bibinfo{author}{Dabringhaus, A.}, \bibinfo{author}{Roland, P.E.},
  \bibinfo{author}{Zilles, K.}, \bibinfo{year}{1996}.
\newblock \bibinfo{title}{Asymmetry in the {Human} {Motor} {Cortex} and
  {Handedness}}.
\newblock \bibinfo{journal}{NeuroImage} \bibinfo{volume}{4},
  \bibinfo{pages}{216--222}.
\newblock \URLprefix
  \url{https://www.sciencedirect.com/science/article/pii/S1053811996900737},
  \DOIprefix\doi{10.1006/nimg.1996.0073}.
%Type = Article
\bibitem[{Auzias et~al.(2014)Auzias, Viellard, Takerkart, Villeneuve, Poinso,
  Fonséca, Girard and Deruelle}]{auzias_atypical_2014}
\bibinfo{author}{Auzias, G.}, \bibinfo{author}{Viellard, M.},
  \bibinfo{author}{Takerkart, S.}, \bibinfo{author}{Villeneuve, N.},
  \bibinfo{author}{Poinso, F.}, \bibinfo{author}{Fonséca, D.D.},
  \bibinfo{author}{Girard, N.}, \bibinfo{author}{Deruelle, C.},
  \bibinfo{year}{2014}.
\newblock \bibinfo{title}{Atypical sulcal anatomy in young children with autism
  spectrum disorder}.
\newblock \bibinfo{journal}{NeuroImage: Clinical} \bibinfo{volume}{4},
  \bibinfo{pages}{593--603}.
\newblock \URLprefix
  \url{https://www.sciencedirect.com/science/article/pii/S2213158214000382},
  \DOIprefix\doi{10.1016/j.nicl.2014.03.008}.
%Type = Article
\bibitem[{Baur et~al.(2020)Baur, Denner, Wiestler, Albarqouni and
  Navab}]{baur_autoencoders_2020}
\bibinfo{author}{Baur, C.}, \bibinfo{author}{Denner, S.},
  \bibinfo{author}{Wiestler, B.}, \bibinfo{author}{Albarqouni, S.},
  \bibinfo{author}{Navab, N.}, \bibinfo{year}{2020}.
\newblock \bibinfo{title}{Autoencoders for {Unsupervised} {Anomaly}
  {Segmentation} in {Brain} {MR} {Images}: {A} {Comparative} {Study}}.
\newblock \bibinfo{journal}{arXiv:2004.03271 [cs, eess]} \URLprefix
  \url{http://arxiv.org/abs/2004.03271}. \bibinfo{note}{arXiv: 2004.03271}.
%Type = Inproceedings
\bibitem[{Behrendt et~al.(2022)Behrendt, Bengs, Rogge, Krüger, Opfer and
  Schlaefer}]{behrendt_unsupervised_2022}
\bibinfo{author}{Behrendt, F.}, \bibinfo{author}{Bengs, M.},
  \bibinfo{author}{Rogge, F.}, \bibinfo{author}{Krüger, J.},
  \bibinfo{author}{Opfer, R.}, \bibinfo{author}{Schlaefer, A.},
  \bibinfo{year}{2022}.
\newblock \bibinfo{title}{Unsupervised {Anomaly} {Detection} in {3D} {Brain}
  {MRI} {Using} {Deep} {Learning} with {Impured} {Training} {Data}}, in:
  \bibinfo{booktitle}{2022 {IEEE} 19th {International} {Symposium} on
  {Biomedical} {Imaging} ({ISBI})}, pp. \bibinfo{pages}{1--4}.
\newblock \DOIprefix\doi{10.1109/ISBI52829.2022.9761443}. \bibinfo{note}{iSSN:
  1945-8452}.
%Type = Article
\bibitem[{Bo et~al.(2015)Bo, Haitao, Yuchun, Zhongyu, Junhai, Xiangtao and
  Shuwei}]{bo_asymmetries_2015}
\bibinfo{author}{Bo, S.}, \bibinfo{author}{Haitao, G.},
  \bibinfo{author}{Yuchun, T.}, \bibinfo{author}{Zhongyu, H.},
  \bibinfo{author}{Junhai, X.}, \bibinfo{author}{Xiangtao, L.},
  \bibinfo{author}{Shuwei, L.}, \bibinfo{year}{2015}.
\newblock \bibinfo{title}{Asymmetries of the central sulcus in young adults:
  {Effects} of gender, age and sulcal pattern}.
\newblock \bibinfo{journal}{International journal of developmental neuroscience
  : the official journal of the International Society for Developmental
  Neuroscience} \bibinfo{volume}{44}.
\newblock \URLprefix \url{https://pubmed.ncbi.nlm.nih.gov/26065979/},
  \DOIprefix\doi{10.1016/j.ijdevneu.2015.06.003}. \bibinfo{note}{publisher: Int
  J Dev Neurosci}.
%Type = Article
\bibitem[{Bodensteiner et~al.(1994)Bodensteiner, Schaefer, Breeding and
  Cowan}]{bodensteiner_hypoplasia_1994}
\bibinfo{author}{Bodensteiner, J.}, \bibinfo{author}{Schaefer, G.},
  \bibinfo{author}{Breeding, L.}, \bibinfo{author}{Cowan, L.},
  \bibinfo{year}{1994}.
\newblock \bibinfo{title}{Hypoplasia of the {Corpus} {Callosum}: {A} {Study} of
  445 {Consecutive} {MRI} {Scans}}.
\newblock \bibinfo{journal}{Journal of Child Neurology} \bibinfo{volume}{9},
  \bibinfo{pages}{47--49}.
\newblock \URLprefix \url{https://doi.org/10.1177/088307389400900111},
  \DOIprefix\doi{10.1177/088307389400900111}. \bibinfo{note}{publisher: SAGE
  Publications Inc}.
%Type = Article
\bibitem[{Boling and Olivier(2004)}]{boling_localization_2004}
\bibinfo{author}{Boling, W.W.}, \bibinfo{author}{Olivier, A.},
  \bibinfo{year}{2004}.
\newblock \bibinfo{title}{Localization of hand sensory function to the pli de
  passage moyen of {Broca}}.
\newblock \bibinfo{journal}{Journal of Neurosurgery} \bibinfo{volume}{101},
  \bibinfo{pages}{278--283}.
\newblock \DOIprefix\doi{10.3171/jns.2004.101.2.0278}.
%Type = Article
\bibitem[{Borne et~al.(2021)Borne, Rivière, Cachia, Roca, Mellerio, Oppenheim
  and Mangin}]{borne_automatic_2021}
\bibinfo{author}{Borne, L.}, \bibinfo{author}{Rivière, D.},
  \bibinfo{author}{Cachia, A.}, \bibinfo{author}{Roca, P.},
  \bibinfo{author}{Mellerio, C.}, \bibinfo{author}{Oppenheim, C.},
  \bibinfo{author}{Mangin, J.F.}, \bibinfo{year}{2021}.
\newblock \bibinfo{title}{Automatic recognition of specific local cortical
  folding patterns}.
\newblock \bibinfo{journal}{NeuroImage} \bibinfo{volume}{238},
  \bibinfo{pages}{118208}.
%Type = Article
\bibitem[{Borne et~al.(2020)Borne, Rivière, Mancip and
  Mangin}]{borne_automatic_2020}
\bibinfo{author}{Borne, L.}, \bibinfo{author}{Rivière, D.},
  \bibinfo{author}{Mancip, M.}, \bibinfo{author}{Mangin, J.F.},
  \bibinfo{year}{2020}.
\newblock \bibinfo{title}{Automatic labeling of cortical sulci using patch- or
  {CNN}-based segmentation techniques combined with bottom-up geometric
  constraints}.
\newblock \bibinfo{journal}{Medical Image Analysis} \bibinfo{volume}{62},
  \bibinfo{pages}{101651}.
\newblock \URLprefix
  \url{http://www.sciencedirect.com/science/article/pii/S1361841520300189},
  \DOIprefix\doi{10.1016/j.media.2020.101651}.
%Type = Article
\bibitem[{Borst et~al.(2014)Borst, Cachia, Vidal, Simon, Pineau, Fischer,
  Poirel, Mangin and Houdé}]{borst_folding_2014}
\bibinfo{author}{Borst, G.}, \bibinfo{author}{Cachia, A.},
  \bibinfo{author}{Vidal, J.}, \bibinfo{author}{Simon, G.},
  \bibinfo{author}{Pineau, A.}, \bibinfo{author}{Fischer, C.},
  \bibinfo{author}{Poirel, N.}, \bibinfo{author}{Mangin, J.F.},
  \bibinfo{author}{Houdé, O.}, \bibinfo{year}{2014}.
\newblock \bibinfo{title}{Folding of the anterior cingulate cortex partially
  explains inhibitory control during childhood: {A} longitudinal study}.
\newblock \bibinfo{journal}{Developmental Cognitive Neuroscience}
  \bibinfo{volume}{9}, \bibinfo{pages}{126--135}.
\newblock \bibinfo{note}{Publisher: Elsevier}.
%Type = Book
\bibitem[{Broca and Pozzi(1888)}]{broca_memoires_1888}
\bibinfo{author}{Broca, P.}, \bibinfo{author}{Pozzi, S.J.},
  \bibinfo{year}{1888}.
\newblock \bibinfo{title}{Mémoires sur le cerveau de l'homme et des primates}.
\newblock \bibinfo{publisher}{C. Reinwald}.
\newblock \bibinfo{note}{Google-Books-ID: d99EAAAAYAAJ}.
%Type = Article
\bibitem[{Bénézit et~al.(2015)Bénézit, Hertz-Pannier, Dehaene-Lambertz,
  Monzalvo, Germanaud, Duclap, Guevara, Mangin, Poupon, Moutard and
  Dubois}]{benezit_organising_2015}
\bibinfo{author}{Bénézit, A.}, \bibinfo{author}{Hertz-Pannier, L.},
  \bibinfo{author}{Dehaene-Lambertz, G.}, \bibinfo{author}{Monzalvo, K.},
  \bibinfo{author}{Germanaud, D.}, \bibinfo{author}{Duclap, D.},
  \bibinfo{author}{Guevara, P.}, \bibinfo{author}{Mangin, J.F.},
  \bibinfo{author}{Poupon, C.}, \bibinfo{author}{Moutard, M.L.},
  \bibinfo{author}{Dubois, J.}, \bibinfo{year}{2015}.
\newblock \bibinfo{title}{Organising white matter in a brain without corpus
  callosum fibres}.
\newblock \bibinfo{journal}{Cortex; a Journal Devoted to the Study of the
  Nervous System and Behavior} \bibinfo{volume}{63}, \bibinfo{pages}{155--171}.
\newblock \DOIprefix\doi{10.1016/j.cortex.2014.08.022}.
%Type = Article
\bibitem[{Cachia et~al.(2016)Cachia, Borst, Tissier, Fisher, Plaze, Gay,
  Rivière, Gogtay, Giedd, Mangin, Houdé and
  Raznahan}]{cachia_longitudinal_2016}
\bibinfo{author}{Cachia, A.}, \bibinfo{author}{Borst, G.},
  \bibinfo{author}{Tissier, C.}, \bibinfo{author}{Fisher, C.},
  \bibinfo{author}{Plaze, M.}, \bibinfo{author}{Gay, O.},
  \bibinfo{author}{Rivière, D.}, \bibinfo{author}{Gogtay, N.},
  \bibinfo{author}{Giedd, J.}, \bibinfo{author}{Mangin, J.F.},
  \bibinfo{author}{Houdé, O.}, \bibinfo{author}{Raznahan, A.},
  \bibinfo{year}{2016}.
\newblock \bibinfo{title}{Longitudinal stability of the folding pattern of the
  anterior cingulate cortex during development}.
\newblock \bibinfo{journal}{Developmental Cognitive Neuroscience}
  \bibinfo{volume}{19}, \bibinfo{pages}{122--127}.
%Type = Article
\bibitem[{Caulo et~al.(2007)Caulo, Briganti, Mattei, Perfetti, Ferretti,
  Romani, Tartaro and Colosimo}]{caulo_new_2007}
\bibinfo{author}{Caulo, M.}, \bibinfo{author}{Briganti, C.},
  \bibinfo{author}{Mattei, P.A.}, \bibinfo{author}{Perfetti, B.},
  \bibinfo{author}{Ferretti, A.}, \bibinfo{author}{Romani, G.L.},
  \bibinfo{author}{Tartaro, A.}, \bibinfo{author}{Colosimo, C.},
  \bibinfo{year}{2007}.
\newblock \bibinfo{title}{New morphologic variants of the hand motor cortex as
  seen with {MR} imaging in a large study population}.
\newblock \bibinfo{journal}{AJNR. American journal of neuroradiology}
  \bibinfo{volume}{28}, \bibinfo{pages}{1480--1485}.
\newblock \DOIprefix\doi{10.3174/ajnr.A0597}.
%Type = Article
\bibitem[{Chalapathy and Chawla(2019)}]{chalapathy_deep_2019}
\bibinfo{author}{Chalapathy, R.}, \bibinfo{author}{Chawla, S.},
  \bibinfo{year}{2019}.
\newblock \bibinfo{title}{Deep {Learning} for {Anomaly} {Detection}: {A}
  {Survey}}.
\newblock \bibinfo{journal}{arXiv:1901.03407 [cs, stat]} \URLprefix
  \url{http://arxiv.org/abs/1901.03407}. \bibinfo{note}{arXiv: 1901.03407}.
%Type = Article
\bibitem[{Davatzikos and Bryan(2002)}]{davatzikos_morphometric_2002}
\bibinfo{author}{Davatzikos, C.}, \bibinfo{author}{Bryan, R.N.},
  \bibinfo{year}{2002}.
\newblock \bibinfo{title}{Morphometric {Analysis} of {Cortical} {Sulci} {Using}
  {Parametric} {Ribbons}: {A} {Study} of the {Central} {Sulcus}:}.
\newblock \bibinfo{journal}{Journal of Computer Assisted Tomography}
  \bibinfo{volume}{26}, \bibinfo{pages}{298--307}.
\newblock \URLprefix \url{http://journals.lww.com/00004728-200203000-00024},
  \DOIprefix\doi{10.1097/00004728-200203000-00024}.
%Type = Article
\bibitem[{Dierker et~al.(2015)Dierker, Feczko, Pruett, Petersen, Schlaggar,
  Constantino, Harwell, Coalson and Van~Essen}]{dierker_analysis_2015}
\bibinfo{author}{Dierker, D.L.}, \bibinfo{author}{Feczko, E.},
  \bibinfo{author}{Pruett, Jr, J.R.}, \bibinfo{author}{Petersen, S.E.},
  \bibinfo{author}{Schlaggar, B.L.}, \bibinfo{author}{Constantino, J.N.},
  \bibinfo{author}{Harwell, J.W.}, \bibinfo{author}{Coalson, T.S.},
  \bibinfo{author}{Van~Essen, D.C.}, \bibinfo{year}{2015}.
\newblock \bibinfo{title}{Analysis of {Cortical} {Shape} in {Children} with
  {Simplex} {Autism}}.
\newblock \bibinfo{journal}{Cerebral Cortex} \bibinfo{volume}{25},
  \bibinfo{pages}{1042--1051}.
\newblock \URLprefix \url{https://doi.org/10.1093/cercor/bht294},
  \DOIprefix\doi{10.1093/cercor/bht294}.
%Type = Article
\bibitem[{Duan et~al.(2019)Duan, Xia, Rekik, Meng, Wu, Wang, Lin, Gilmore, Shen
  and Li}]{duan_exploring_2019}
\bibinfo{author}{Duan, D.}, \bibinfo{author}{Xia, S.}, \bibinfo{author}{Rekik,
  I.}, \bibinfo{author}{Meng, Y.}, \bibinfo{author}{Wu, Z.},
  \bibinfo{author}{Wang, L.}, \bibinfo{author}{Lin, W.},
  \bibinfo{author}{Gilmore, J.H.}, \bibinfo{author}{Shen, D.},
  \bibinfo{author}{Li, G.}, \bibinfo{year}{2019}.
\newblock \bibinfo{title}{Exploring folding patterns of infant cerebral cortex
  based on multi-view curvature features: {Methods} and applications}.
\newblock \bibinfo{journal}{NeuroImage} \bibinfo{volume}{185},
  \bibinfo{pages}{575--592}.
%Type = Article
\bibitem[{Essen et~al.(2006)Essen, Dierker, Snyder, Raichle, Reiss and
  Korenberg}]{essen_symmetry_2006}
\bibinfo{author}{Essen, D.C.V.}, \bibinfo{author}{Dierker, D.},
  \bibinfo{author}{Snyder, A.Z.}, \bibinfo{author}{Raichle, M.E.},
  \bibinfo{author}{Reiss, A.L.}, \bibinfo{author}{Korenberg, J.},
  \bibinfo{year}{2006}.
\newblock \bibinfo{title}{Symmetry of {Cortical} {Folding} {Abnormalities} in
  {Williams} {Syndrome} {Revealed} by {Surface}-{Based} {Analyses}}.
\newblock \bibinfo{journal}{Journal of Neuroscience} \bibinfo{volume}{26},
  \bibinfo{pages}{5470--5483}.
\newblock \URLprefix \url{https://www.jneurosci.org/content/26/20/5470},
  \DOIprefix\doi{10.1523/JNEUROSCI.4154-05.2006}. \bibinfo{note}{publisher:
  Society for Neuroscience Section: Articles}.
%Type = Article
\bibitem[{Fernando et~al.(2022)Fernando, Gammulle, Denman, Sridharan and
  Fookes}]{fernando_deep_2022}
\bibinfo{author}{Fernando, T.}, \bibinfo{author}{Gammulle, H.},
  \bibinfo{author}{Denman, S.}, \bibinfo{author}{Sridharan, S.},
  \bibinfo{author}{Fookes, C.}, \bibinfo{year}{2022}.
\newblock \bibinfo{title}{Deep {Learning} for {Medical} {Anomaly} {Detection}
  – {A} {Survey}}.
\newblock \bibinfo{journal}{ACM Computing Surveys} \bibinfo{volume}{54},
  \bibinfo{pages}{1--37}.
\newblock \URLprefix \url{https://dl.acm.org/doi/10.1145/3464423},
  \DOIprefix\doi{10.1145/3464423}.
%Type = Article
\bibitem[{Fernández et~al.(2016)Fernández, Llinares-Benadero and
  Borrell}]{fernandez_cerebral_2016}
\bibinfo{author}{Fernández, V.}, \bibinfo{author}{Llinares-Benadero, C.},
  \bibinfo{author}{Borrell, V.}, \bibinfo{year}{2016}.
\newblock \bibinfo{title}{Cerebral cortex expansion and folding: what have we
  learned?}
\newblock \bibinfo{journal}{The EMBO Journal} \bibinfo{volume}{35},
  \bibinfo{pages}{1021--1044}.
\newblock \URLprefix
  \url{https://www.embopress.org/doi/full/10.15252/embj.201593701},
  \DOIprefix\doi{10.15252/embj.201593701}. \bibinfo{note}{publisher: John Wiley
  \& Sons, Ltd}.
%Type = Inproceedings
\bibitem[{Foubet et~al.(2022)Foubet, Sun, Hopkins and
  Mangin}]{foubet_comparison_2022}
\bibinfo{author}{Foubet, O.}, \bibinfo{author}{Sun, Z.Y.},
  \bibinfo{author}{Hopkins, W.}, \bibinfo{author}{Mangin, J.F.},
  \bibinfo{year}{2022}.
\newblock \bibinfo{title}{Comparison of the shape of the {Central} {Sulcus} in
  {Hominids}}, in: \bibinfo{booktitle}{Organisation for {Human} {Brain}
  {Mapping}}, \bibinfo{address}{Glasgow, United Kingdom}.
%Type = Article
\bibitem[{Germann et~al.(2020)Germann, Chakravarty, Collins and
  Petrides}]{germann_tight_2020}
\bibinfo{author}{Germann, J.}, \bibinfo{author}{Chakravarty, M.M.},
  \bibinfo{author}{Collins, D.L.}, \bibinfo{author}{Petrides, M.},
  \bibinfo{year}{2020}.
\newblock \bibinfo{title}{Tight {Coupling} between {Morphological} {Features}
  of the {Central} {Sulcus} and {Somatomotor} {Body} {Representations}: {A}
  {Combined} {Anatomical} and {Functional} {MRI} {Study}}.
\newblock \bibinfo{journal}{Cerebral Cortex} \bibinfo{volume}{30},
  \bibinfo{pages}{1843--1854}.
\newblock \URLprefix \url{https://doi.org/10.1093/cercor/bhz208},
  \DOIprefix\doi{10.1093/cercor/bhz208}.
%Type = Article
\bibitem[{Gervais et~al.(2004)Gervais, Belin, Boddaert, Leboyer, Coez, Sfaello,
  Barthélémy, Brunelle, Samson and Zilbovicius}]{gervais_abnormal_2004}
\bibinfo{author}{Gervais, H.}, \bibinfo{author}{Belin, P.},
  \bibinfo{author}{Boddaert, N.}, \bibinfo{author}{Leboyer, M.},
  \bibinfo{author}{Coez, A.}, \bibinfo{author}{Sfaello, I.},
  \bibinfo{author}{Barthélémy, C.}, \bibinfo{author}{Brunelle, F.},
  \bibinfo{author}{Samson, Y.}, \bibinfo{author}{Zilbovicius, M.},
  \bibinfo{year}{2004}.
\newblock \bibinfo{title}{Abnormal cortical voice processing in autism}.
\newblock \bibinfo{journal}{Nature Neuroscience} \bibinfo{volume}{7},
  \bibinfo{pages}{801--802}.
\newblock \DOIprefix\doi{10.1038/nn1291}.
%Type = Inproceedings
\bibitem[{Guillon et~al.(2021)Guillon, Cagna, Dufumier, Chavas, Rivière and
  Mangin}]{guillon_detection_2021}
\bibinfo{author}{Guillon, L.}, \bibinfo{author}{Cagna, B.},
  \bibinfo{author}{Dufumier, B.}, \bibinfo{author}{Chavas, J.},
  \bibinfo{author}{Rivière, D.}, \bibinfo{author}{Mangin, J.F.},
  \bibinfo{year}{2021}.
\newblock \bibinfo{title}{Detection of {Abnormal} {Folding} {Patterns} with
  {Unsupervised} {Deep} {Generative} {Models}}, in:
  \bibinfo{editor}{Abdulkadir, A.}, \bibinfo{editor}{Kia, S.M.},
  \bibinfo{editor}{Habes, M.}, \bibinfo{editor}{Kumar, V.},
  \bibinfo{editor}{Rondina, J.M.}, \bibinfo{editor}{Tax, C.},
  \bibinfo{editor}{Wolfers, T.} (Eds.), \bibinfo{booktitle}{Machine {Learning}
  in {Clinical} {Neuroimaging}}, \bibinfo{publisher}{Springer International
  Publishing}, \bibinfo{address}{Cham}. pp. \bibinfo{pages}{63--72}.
\newblock \DOIprefix\doi{10.1007/978-3-030-87586-2_7}.
%Type = Inproceedings
\bibitem[{Guillon et~al.(2022)Guillon, Chavas, Pascucci, Dufumier, Rivière and
  Mangin}]{guillon_unsupervised_2022}
\bibinfo{author}{Guillon, L.}, \bibinfo{author}{Chavas, J.},
  \bibinfo{author}{Pascucci, M.}, \bibinfo{author}{Dufumier, B.},
  \bibinfo{author}{Rivière, D.}, \bibinfo{author}{Mangin, J.F.},
  \bibinfo{year}{2022}.
\newblock \bibinfo{title}{Unsupervised {Representation} {Learning}
  of {Cingulate} {Cortical} {Folding} {Patterns}}, in: \bibinfo{editor}{Wang,
  L.}, \bibinfo{editor}{Dou, Q.}, \bibinfo{editor}{Fletcher, P.T.},
  \bibinfo{editor}{Speidel, S.}, \bibinfo{editor}{Li, S.} (Eds.),
  \bibinfo{booktitle}{Medical {Image} {Computing} and {Computer} {Assisted}
  {Intervention} – {MICCAI} 2022}, \bibinfo{publisher}{Springer Nature
  Switzerland}, \bibinfo{address}{Cham}. pp. \bibinfo{pages}{77--87}.
\newblock \DOIprefix\doi{10.1007/978-3-031-16431-6_8}.
%Type = Inproceedings
\bibitem[{Higgins et~al.(2017)Higgins, Matthey, Pal, Burgess, Glorot,
  Botvinick, Mohamed and Lerchner}]{higgins_beta-vae_2017}
\bibinfo{author}{Higgins, I.}, \bibinfo{author}{Matthey, L.},
  \bibinfo{author}{Pal, A.}, \bibinfo{author}{Burgess, C.P.},
  \bibinfo{author}{Glorot, X.}, \bibinfo{author}{Botvinick, M.},
  \bibinfo{author}{Mohamed, S.}, \bibinfo{author}{Lerchner, A.},
  \bibinfo{year}{2017}.
\newblock \bibinfo{title}{beta-{VAE}: {Learning} {Basic} {Visual} {Concepts}
  with a {Constrained} {Variational} {Framework}}, in:
  \bibinfo{booktitle}{{ICLR}}.
%Type = Article
\bibitem[{Hopkins et~al.(2014)Hopkins, Meguerditchian, Coulon, Bogart, Mangin,
  Sherwood, Grabowski, Bennett, Pierre, Fears, Woods, Hof and
  Vauclair}]{hopkins_evolution_2014}
\bibinfo{author}{Hopkins, W.D.}, \bibinfo{author}{Meguerditchian, A.},
  \bibinfo{author}{Coulon, O.}, \bibinfo{author}{Bogart, S.},
  \bibinfo{author}{Mangin, J.F.}, \bibinfo{author}{Sherwood, C.C.},
  \bibinfo{author}{Grabowski, M.W.}, \bibinfo{author}{Bennett, A.J.},
  \bibinfo{author}{Pierre, P.J.}, \bibinfo{author}{Fears, S.},
  \bibinfo{author}{Woods, R.}, \bibinfo{author}{Hof, P.R.},
  \bibinfo{author}{Vauclair, J.}, \bibinfo{year}{2014}.
\newblock \bibinfo{title}{Evolution of the central sulcus morphology in
  primates}.
\newblock \bibinfo{journal}{Brain, Behavior and Evolution}
  \bibinfo{volume}{84}, \bibinfo{pages}{19--30}.
\newblock \DOIprefix\doi{10.1159/000362431}.
%Type = Article
\bibitem[{Hotier et~al.(2017)Hotier, Leroy, Boisgontier, Laidi, Mangin,
  Delorme, Bolognani, Czech, Bouquet, Toledano, Bouvard, Petit, Mishchenko,
  d'Albis, Gras, Gaman, Scheid, Leboyer, Zalla and
  Houenou}]{hotier_social_2017}
\bibinfo{author}{Hotier, S.}, \bibinfo{author}{Leroy, F.},
  \bibinfo{author}{Boisgontier, J.}, \bibinfo{author}{Laidi, C.},
  \bibinfo{author}{Mangin, J.F.}, \bibinfo{author}{Delorme, R.},
  \bibinfo{author}{Bolognani, F.}, \bibinfo{author}{Czech, C.},
  \bibinfo{author}{Bouquet, C.}, \bibinfo{author}{Toledano, E.},
  \bibinfo{author}{Bouvard, M.}, \bibinfo{author}{Petit, J.},
  \bibinfo{author}{Mishchenko, M.}, \bibinfo{author}{d'Albis, M.A.},
  \bibinfo{author}{Gras, D.}, \bibinfo{author}{Gaman, A.},
  \bibinfo{author}{Scheid, I.}, \bibinfo{author}{Leboyer, M.},
  \bibinfo{author}{Zalla, T.}, \bibinfo{author}{Houenou, J.},
  \bibinfo{year}{2017}.
\newblock \bibinfo{title}{Social cognition in autism is associated with the
  neurodevelopment of the posterior superior temporal sulcus}.
\newblock \bibinfo{journal}{Acta Psychiatrica Scandinavica}
  \bibinfo{volume}{136}, \bibinfo{pages}{517--525}.
\newblock \DOIprefix\doi{10.1111/acps.12814}.
%Type = Article
\bibitem[{Im et~al.(2011)Im, Pienaar, Lee, Seong, Choi, Lee and
  Grant}]{im_quantitative_2011}
\bibinfo{author}{Im, K.}, \bibinfo{author}{Pienaar, R.}, \bibinfo{author}{Lee,
  J.M.}, \bibinfo{author}{Seong, J.K.}, \bibinfo{author}{Choi, Y.Y.},
  \bibinfo{author}{Lee, K.H.}, \bibinfo{author}{Grant, P.E.},
  \bibinfo{year}{2011}.
\newblock \bibinfo{title}{Quantitative comparison and analysis of sulcal
  patterns using sulcal graph matching: a twin study}.
\newblock \bibinfo{journal}{NeuroImage} \bibinfo{volume}{57},
  \bibinfo{pages}{1077--1086}.
\newblock \DOIprefix\doi{10.1016/j.neuroimage.2011.04.062}.
%Type = Article
\bibitem[{Jin et~al.(2018)Jin, Zhang, Shaw, Sachdev and
  Cherbuin}]{jin_relationship_2018}
\bibinfo{author}{Jin, K.}, \bibinfo{author}{Zhang, T.}, \bibinfo{author}{Shaw,
  M.}, \bibinfo{author}{Sachdev, P.}, \bibinfo{author}{Cherbuin, N.},
  \bibinfo{year}{2018}.
\newblock \bibinfo{title}{Relationship {Between} {Sulcal} {Characteristics} and
  {Brain} {Aging}}.
\newblock \bibinfo{journal}{Frontiers in Aging Neuroscience}
  \bibinfo{volume}{10}.
\newblock \URLprefix
  \url{https://www.frontiersin.org/articles/10.3389/fnagi.2018.00339}.
%Type = Article
\bibitem[{Kingma and Welling(2014)}]{kingma_auto-encoding_2014}
\bibinfo{author}{Kingma, D.P.}, \bibinfo{author}{Welling, M.},
  \bibinfo{year}{2014}.
\newblock \bibinfo{title}{Auto-{Encoding} {Variational} {Bayes}}.
\newblock \bibinfo{journal}{arXiv:1312.6114 [cs, stat]} \URLprefix
  \url{http://arxiv.org/abs/1312.6114}. \bibinfo{note}{arXiv: 1312.6114}.
%Type = Article
\bibitem[{Kochunov et~al.(2005)Kochunov, Mangin, Coyle, Lancaster, Thompson,
  Rivière, Cointepas, Régis, Schlosser, Royall, Zilles, Mazziotta, Toga and
  Fox}]{kochunov_age-related_2005}
\bibinfo{author}{Kochunov, P.}, \bibinfo{author}{Mangin, J.F.},
  \bibinfo{author}{Coyle, T.}, \bibinfo{author}{Lancaster, J.},
  \bibinfo{author}{Thompson, P.}, \bibinfo{author}{Rivière, D.},
  \bibinfo{author}{Cointepas, Y.}, \bibinfo{author}{Régis, J.},
  \bibinfo{author}{Schlosser, A.}, \bibinfo{author}{Royall, D.R.},
  \bibinfo{author}{Zilles, K.}, \bibinfo{author}{Mazziotta, J.},
  \bibinfo{author}{Toga, A.}, \bibinfo{author}{Fox, P.T.},
  \bibinfo{year}{2005}.
\newblock \bibinfo{title}{Age-related morphology trends of cortical sulci}.
\newblock \bibinfo{journal}{Human Brain Mapping} \bibinfo{volume}{26},
  \bibinfo{pages}{210--220}.
\newblock \URLprefix
  \url{https://onlinelibrary.wiley.com/doi/abs/10.1002/hbm.20198},
  \DOIprefix\doi{10.1002/hbm.20198}. \bibinfo{note}{\_eprint:
  https://onlinelibrary.wiley.com/doi/pdf/10.1002/hbm.20198}.
%Type = Article
\bibitem[{Levitt et~al.(2003)Levitt, Blanton, Smalley, Thompson, Guthrie,
  McCracken, Sadoun, Heinichen and Toga}]{levitt_cortical_2003}
\bibinfo{author}{Levitt, J.G.}, \bibinfo{author}{Blanton, R.E.},
  \bibinfo{author}{Smalley, S.}, \bibinfo{author}{Thompson, P.},
  \bibinfo{author}{Guthrie, D.}, \bibinfo{author}{McCracken, J.T.},
  \bibinfo{author}{Sadoun, T.}, \bibinfo{author}{Heinichen, L.},
  \bibinfo{author}{Toga, A.W.}, \bibinfo{year}{2003}.
\newblock \bibinfo{title}{Cortical {Sulcal} {Maps} in {Autism}}.
\newblock \bibinfo{journal}{Cerebral Cortex} \bibinfo{volume}{13},
  \bibinfo{pages}{728--735}.
\newblock \URLprefix \url{https://doi.org/10.1093/cercor/13.7.728},
  \DOIprefix\doi{10.1093/cercor/13.7.728}.
%Type = Inproceedings
\bibitem[{Liu et~al.(2008)Liu, Ting and Zhou}]{liu_isolation_2008}
\bibinfo{author}{Liu, F.T.}, \bibinfo{author}{Ting, K.M.},
  \bibinfo{author}{Zhou, Z.H.}, \bibinfo{year}{2008}.
\newblock \bibinfo{title}{Isolation {Forest}}, in: \bibinfo{booktitle}{2008
  {Eighth} {IEEE} {International} {Conference} on {Data} {Mining}}, pp.
  \bibinfo{pages}{413--422}.
\newblock \DOIprefix\doi{10.1109/ICDM.2008.17}. \bibinfo{note}{iSSN:
  2374-8486}.
%Type = Article
\bibitem[{Lohmann et~al.(2008)Lohmann, von Cramon and
  Colchester}]{lohmann_deep_2008}
\bibinfo{author}{Lohmann, G.}, \bibinfo{author}{von Cramon, D.Y.},
  \bibinfo{author}{Colchester, A.C.F.}, \bibinfo{year}{2008}.
\newblock \bibinfo{title}{Deep sulcal landmarks provide an organizing framework
  for human cortical folding}.
\newblock \bibinfo{journal}{Cerebral Cortex (New York, N.Y.: 1991)}
  \bibinfo{volume}{18}, \bibinfo{pages}{1415--1420}.
\newblock \DOIprefix\doi{10.1093/cercor/bhm174}.
%Type = Article
\bibitem[{Mangin et~al.(1995)Mangin, Frouin, Bloch, Rigis and
  Lopez-Krahe}]{mangin_3d_1995}
\bibinfo{author}{Mangin, J.F.}, \bibinfo{author}{Frouin, V.},
  \bibinfo{author}{Bloch, I.}, \bibinfo{author}{Rigis, J.},
  \bibinfo{author}{Lopez-Krahe, J.}, \bibinfo{year}{1995}.
\newblock \bibinfo{title}{From {3D} magnetic resonance images to structural
  representations of the cortex topography using topology preserving
  deformations}.
\newblock \bibinfo{journal}{Journal of Mathematical imaging and Vision}
  \bibinfo{volume}{5}, \bibinfo{pages}{297--318}.
%Type = Article
\bibitem[{Mangin et~al.(2019)Mangin, Le~Guen, Labra, Grigis, Frouin, Guevara,
  Fischer, Rivière, Hopkins, Régis and Sun}]{mangin_plis_2019}
\bibinfo{author}{Mangin, J.F.}, \bibinfo{author}{Le~Guen, Y.},
  \bibinfo{author}{Labra, N.}, \bibinfo{author}{Grigis, A.},
  \bibinfo{author}{Frouin, V.}, \bibinfo{author}{Guevara, M.},
  \bibinfo{author}{Fischer, C.}, \bibinfo{author}{Rivière, D.},
  \bibinfo{author}{Hopkins, W.D.}, \bibinfo{author}{Régis, J.},
  \bibinfo{author}{Sun, Z.Y.}, \bibinfo{year}{2019}.
\newblock \bibinfo{title}{“{Plis} de passage” {Deserve} a {Role} in
  {Models} of the {Cortical} {Folding} {Process}}.
\newblock \bibinfo{journal}{Brain Topography} \bibinfo{volume}{32},
  \bibinfo{pages}{1035--1048}.
\newblock \URLprefix \url{http://link.springer.com/10.1007/s10548-019-00734-8},
  \DOIprefix\doi{10.1007/s10548-019-00734-8}.
%Type = Misc
\bibitem[{McInnes et~al.(2020)McInnes, Healy and Melville}]{mcinnes_umap_2020}
\bibinfo{author}{McInnes, L.}, \bibinfo{author}{Healy, J.},
  \bibinfo{author}{Melville, J.}, \bibinfo{year}{2020}.
\newblock \bibinfo{title}{{UMAP}: {Uniform} {Manifold} {Approximation} and
  {Projection} for {Dimension} {Reduction}}.
\newblock \URLprefix \url{http://arxiv.org/abs/1802.03426}.
  \bibinfo{note}{arXiv:1802.03426 [cs, stat]}.
%Type = Article
\bibitem[{McInnes et~al.(2018)McInnes, Healy, Saul and
  Grossberger}]{mcinnes_umap_2018}
\bibinfo{author}{McInnes, L.}, \bibinfo{author}{Healy, J.},
  \bibinfo{author}{Saul, N.}, \bibinfo{author}{Grossberger, L.},
  \bibinfo{year}{2018}.
\newblock \bibinfo{title}{{UMAP}: {Uniform} {Manifold} {Approximation} and
  {Projection}}.
\newblock \bibinfo{journal}{The Journal of Open Source Software}
  \bibinfo{volume}{3}, \bibinfo{pages}{861}.
%Type = Inproceedings
\bibitem[{Meissen et~al.(2022)Meissen, Wiestler, Kaissis and
  Rueckert}]{meissen_pitfalls_2022}
\bibinfo{author}{Meissen, F.}, \bibinfo{author}{Wiestler, B.},
  \bibinfo{author}{Kaissis, G.}, \bibinfo{author}{Rueckert, D.},
  \bibinfo{year}{2022}.
\newblock \bibinfo{title}{On the {Pitfalls} of {Using} the {Residual} as
  {Anomaly} {Score}}, in: \bibinfo{booktitle}{Medical {Imaging} with {Deep}
  {Learning}}.
\newblock \URLprefix \url{https://openreview.net/forum?id=ZsoHLeupa1D}.
%Type = Article
\bibitem[{Mellerio et~al.(2014)Mellerio, Roca, Chassoux, Danière, Cachia,
  Lion, Naggara, Devaux, Meder and Oppenheim}]{mellerio_power_2014}
\bibinfo{author}{Mellerio, C.}, \bibinfo{author}{Roca, P.},
  \bibinfo{author}{Chassoux, F.}, \bibinfo{author}{Danière, F.},
  \bibinfo{author}{Cachia, A.}, \bibinfo{author}{Lion, S.},
  \bibinfo{author}{Naggara, O.}, \bibinfo{author}{Devaux, B.},
  \bibinfo{author}{Meder, J.F.}, \bibinfo{author}{Oppenheim, C.},
  \bibinfo{year}{2014}.
\newblock \bibinfo{title}{The {Power} {Button} {Sign}: {A} {Newly} {Described}
  {Central} {Sulcal} {Pattern} on {Surface} {Rendering} {MR} {Images} of {Type}
  2 {Focal} {Cortical} {Dysplasia}}.
\newblock \bibinfo{journal}{Radiology} \bibinfo{volume}{274},
  \bibinfo{pages}{500--507}.
\newblock \URLprefix
  \url{https://pubs.rsna.org/doi/full/10.1148/radiol.14140773},
  \DOIprefix\doi{10.1148/radiol.14140773}. \bibinfo{note}{publisher:
  Radiological Society of North America}.
%Type = Article
\bibitem[{Meng et~al.(2018)Meng, Li, Wang, Lin, Gilmore and
  Shen}]{meng_discovering_2018}
\bibinfo{author}{Meng, Y.}, \bibinfo{author}{Li, G.}, \bibinfo{author}{Wang,
  L.}, \bibinfo{author}{Lin, W.}, \bibinfo{author}{Gilmore, J.H.},
  \bibinfo{author}{Shen, D.}, \bibinfo{year}{2018}.
\newblock \bibinfo{title}{Discovering cortical sulcal folding patterns in
  neonates using large‐scale dataset}.
\newblock \bibinfo{journal}{Human Brain Mapping} \bibinfo{volume}{39},
  \bibinfo{pages}{3625--3635}.
%Type = Article
\bibitem[{Nordahl et~al.(2007)Nordahl, Dierker, Mostafavi, Schumann, Rivera,
  Amaral and Van~Essen}]{nordahl_cortical_2007}
\bibinfo{author}{Nordahl, C.W.}, \bibinfo{author}{Dierker, D.},
  \bibinfo{author}{Mostafavi, I.}, \bibinfo{author}{Schumann, C.M.},
  \bibinfo{author}{Rivera, S.M.}, \bibinfo{author}{Amaral, D.G.},
  \bibinfo{author}{Van~Essen, D.C.}, \bibinfo{year}{2007}.
\newblock \bibinfo{title}{Cortical {Folding} {Abnormalities} in {Autism}
  {Revealed} by {Surface}-{Based} {Morphometry}}.
\newblock \bibinfo{journal}{The Journal of Neuroscience} \bibinfo{volume}{27},
  \bibinfo{pages}{11725--11735}.
\newblock \URLprefix
  \url{https://www.ncbi.nlm.nih.gov/pmc/articles/PMC6673212/},
  \DOIprefix\doi{10.1523/JNEUROSCI.0777-07.2007}.
%Type = Book
\bibitem[{Ono et~al.(1990)Ono, Kubik and Abarnathey}]{ono_atlas_1990}
\bibinfo{author}{Ono, M.}, \bibinfo{author}{Kubik, S.},
  \bibinfo{author}{Abarnathey, C.D.}, \bibinfo{year}{1990}.
\newblock \bibinfo{title}{Atlas of the {Cerebral} {Sulci}}.
\newblock \bibinfo{edition}{1er édition} ed.,
  \bibinfo{publisher}{Thieme-Stratton Corp}, \bibinfo{address}{Stuttgart : New
  York}.
%Type = Inproceedings
\bibitem[{van~den Oord et~al.(2017)van~den Oord, Vinyals and
  kavukcuoglu}]{van_den_oord_neural_2017}
\bibinfo{author}{van~den Oord, A.}, \bibinfo{author}{Vinyals, O.},
  \bibinfo{author}{kavukcuoglu, k.}, \bibinfo{year}{2017}.
\newblock \bibinfo{title}{Neural {Discrete} {Representation} {Learning}}, in:
  \bibinfo{booktitle}{Advances in {Neural} {Information} {Processing}
  {Systems}}, \bibinfo{publisher}{Curran Associates, Inc.}
\newblock \URLprefix
  \url{https://papers.nips.cc/paper/2017/hash/7a98af17e63a0ac09ce2e96d03992fbc-Abstract.html}.
%Type = Article
\bibitem[{Pedregosa et~al.(2011)Pedregosa, Varoquaux, Gramfort, Michel,
  Thirion, Grisel, Blondel, Prettenhofer, Weiss, Dubourg, Vanderplas, Passos,
  Cournapeau, Brucher, Perrot and Duchesnay}]{pedregosa_scikit-learn_2011}
\bibinfo{author}{Pedregosa, F.}, \bibinfo{author}{Varoquaux, G.},
  \bibinfo{author}{Gramfort, A.}, \bibinfo{author}{Michel, V.},
  \bibinfo{author}{Thirion, B.}, \bibinfo{author}{Grisel, O.},
  \bibinfo{author}{Blondel, M.}, \bibinfo{author}{Prettenhofer, P.},
  \bibinfo{author}{Weiss, R.}, \bibinfo{author}{Dubourg, V.},
  \bibinfo{author}{Vanderplas, J.}, \bibinfo{author}{Passos, A.},
  \bibinfo{author}{Cournapeau, D.}, \bibinfo{author}{Brucher, M.},
  \bibinfo{author}{Perrot, M.}, \bibinfo{author}{Duchesnay, E.},
  \bibinfo{year}{2011}.
\newblock \bibinfo{title}{Scikit-learn: {Machine} {Learning} in {Python}}.
\newblock \bibinfo{journal}{Journal of Machine Learning Research}
  \bibinfo{volume}{12}, \bibinfo{pages}{2825--2830}.
\newblock \URLprefix \url{http://jmlr.org/papers/v12/pedregosa11a.html}.
%Type = Article
\bibitem[{Pinaya et~al.(2018)Pinaya, Mechelli and Sato}]{pinaya_using_2018}
\bibinfo{author}{Pinaya, W.H.L.}, \bibinfo{author}{Mechelli, A.},
  \bibinfo{author}{Sato, J.R.}, \bibinfo{year}{2018}.
\newblock \bibinfo{title}{Using deep autoencoders to identify abnormal brain
  structural patterns in neuropsychiatric disorders: {A} large‐scale
  multi‐sample study}.
\newblock \bibinfo{journal}{Human Brain Mapping} \bibinfo{volume}{40},
  \bibinfo{pages}{944--954}.
\newblock \URLprefix
  \url{https://www.ncbi.nlm.nih.gov/pmc/articles/PMC6492107/},
  \DOIprefix\doi{10.1002/hbm.24423}.
%Type = Article
\bibitem[{Provost et~al.(2003)Provost, Bartrés-Faz, Paillère-Martinot,
  Artiges, Pappata, Recasens, Pérez-Gómez, Bernardo, Baeza, Bayle and
  Martinot}]{provost_paracingulate_2003}
\bibinfo{author}{Provost, J.B.L.}, \bibinfo{author}{Bartrés-Faz, D.},
  \bibinfo{author}{Paillère-Martinot, M.L.}, \bibinfo{author}{Artiges, E.},
  \bibinfo{author}{Pappata, S.}, \bibinfo{author}{Recasens, C.},
  \bibinfo{author}{Pérez-Gómez, M.}, \bibinfo{author}{Bernardo, M.},
  \bibinfo{author}{Baeza, I.}, \bibinfo{author}{Bayle, F.},
  \bibinfo{author}{Martinot, J.L.}, \bibinfo{year}{2003}.
\newblock \bibinfo{title}{Paracingulate sulcus morphology in men with
  early-onset schizophrenia}.
\newblock \bibinfo{journal}{British Journal of Psychiatry}
  \bibinfo{volume}{182}, \bibinfo{pages}{228--232}.
%Type = Misc
\bibitem[{Qiu et~al.(2022)Qiu, Li, Kloft, Rudolph and Mandt}]{qiu_latent_2022}
\bibinfo{author}{Qiu, C.}, \bibinfo{author}{Li, A.}, \bibinfo{author}{Kloft,
  M.}, \bibinfo{author}{Rudolph, M.}, \bibinfo{author}{Mandt, S.},
  \bibinfo{year}{2022}.
\newblock \bibinfo{title}{Latent {Outlier} {Exposure} for {Anomaly} {Detection}
  with {Contaminated} {Data}}.
\newblock \URLprefix \url{http://arxiv.org/abs/2202.08088}.
  \bibinfo{note}{number: arXiv:2202.08088 arXiv:2202.08088 [cs]}.
%Type = Article
\bibitem[{Rivière et~al.(2002)Rivière, Mangin, Papadopoulos-Orfanos,
  Martinez, Frouin and Régis}]{riviere_automatic_2002}
\bibinfo{author}{Rivière, D.}, \bibinfo{author}{Mangin, J.F.},
  \bibinfo{author}{Papadopoulos-Orfanos, D.}, \bibinfo{author}{Martinez, J.M.},
  \bibinfo{author}{Frouin, V.}, \bibinfo{author}{Régis, J.},
  \bibinfo{year}{2002}.
\newblock \bibinfo{title}{Automatic recognition of cortical sulci of the human
  brain using a congregation of neural networks}.
\newblock \bibinfo{journal}{Medical Image Analysis} \bibinfo{volume}{6},
  \bibinfo{pages}{77--92}.
\newblock \URLprefix
  \url{https://linkinghub.elsevier.com/retrieve/pii/S136184150200052X},
  \DOIprefix\doi{10.1016/S1361-8415(02)00052-X}.
%Type = Article
\bibitem[{Roy et~al.(2020)Roy, McMillen, Beiler, Snyder, Patti and
  Troiani}]{roy_pipeline_2020}
\bibinfo{author}{Roy, A.}, \bibinfo{author}{McMillen, T.},
  \bibinfo{author}{Beiler, D.L.}, \bibinfo{author}{Snyder, W.},
  \bibinfo{author}{Patti, M.}, \bibinfo{author}{Troiani, V.},
  \bibinfo{year}{2020}.
\newblock \bibinfo{title}{A pipeline to characterize local cortical folds by
  mapping them to human-interpretable shapes}.
\newblock \bibinfo{journal}{bioRxiv} \URLprefix
  \url{https://www.biorxiv.org/content/early/2020/11/26/2020.11.25.388785},
  \DOIprefix\doi{10.1101/2020.11.25.388785}. \bibinfo{note}{publisher: Cold
  Spring Harbor Laboratory \_eprint:
  https://www.biorxiv.org/content/early/2020/11/26/2020.11.25.388785.full.pdf}.
%Type = Inproceedings
\bibitem[{Ruff et~al.(2018)Ruff, Vandermeulen, Goernitz, Deecke, Siddiqui,
  Binder, Müller and Kloft}]{ruff_deep_2018}
\bibinfo{author}{Ruff, L.}, \bibinfo{author}{Vandermeulen, R.},
  \bibinfo{author}{Goernitz, N.}, \bibinfo{author}{Deecke, L.},
  \bibinfo{author}{Siddiqui, S.A.}, \bibinfo{author}{Binder, A.},
  \bibinfo{author}{Müller, E.}, \bibinfo{author}{Kloft, M.},
  \bibinfo{year}{2018}.
\newblock \bibinfo{title}{Deep {One}-{Class} {Classification}}, in:
  \bibinfo{booktitle}{Proceedings of the 35th {International} {Conference} on
  {Machine} {Learning}}, \bibinfo{publisher}{PMLR}. pp.
  \bibinfo{pages}{4393--4402}.
\newblock \URLprefix \url{https://proceedings.mlr.press/v80/ruff18a.html}.
  \bibinfo{note}{iSSN: 2640-3498}.
%Type = Article
\bibitem[{Schlegl et~al.(2019)Schlegl, Seeböck, Waldstein, Langs and
  Schmidt-Erfurth}]{schlegl_f-anogan_2019}
\bibinfo{author}{Schlegl, T.}, \bibinfo{author}{Seeböck, P.},
  \bibinfo{author}{Waldstein, S.M.}, \bibinfo{author}{Langs, G.},
  \bibinfo{author}{Schmidt-Erfurth, U.}, \bibinfo{year}{2019}.
\newblock \bibinfo{title}{f-{AnoGAN}: {Fast} unsupervised anomaly detection
  with generative adversarial networks}.
\newblock \bibinfo{journal}{Medical Image Analysis} \bibinfo{volume}{54},
  \bibinfo{pages}{30--44}.
\newblock \URLprefix
  \url{https://linkinghub.elsevier.com/retrieve/pii/S1361841518302640},
  \DOIprefix\doi{10.1016/j.media.2019.01.010}.
%Type = Article
\bibitem[{Schlegl et~al.(2017)Schlegl, Seeböck, Waldstein, Schmidt-Erfurth and
  Langs}]{schlegl_unsupervised_2017}
\bibinfo{author}{Schlegl, T.}, \bibinfo{author}{Seeböck, P.},
  \bibinfo{author}{Waldstein, S.M.}, \bibinfo{author}{Schmidt-Erfurth, U.},
  \bibinfo{author}{Langs, G.}, \bibinfo{year}{2017}.
\newblock \bibinfo{title}{Unsupervised {Anomaly} {Detection} with {Generative}
  {Adversarial} {Networks} to {Guide} {Marker} {Discovery}}.
\newblock \bibinfo{journal}{arXiv:1703.05921 [cs]} \URLprefix
  \url{http://arxiv.org/abs/1703.05921}. \bibinfo{note}{arXiv: 1703.05921}.
%Type = Article
\bibitem[{Schölkopf et~al.(2001)Schölkopf, Platt, Shawe-Taylor, Smola and
  Williamson}]{scholkopf_estimating_2001}
\bibinfo{author}{Schölkopf, B.}, \bibinfo{author}{Platt, J.C.},
  \bibinfo{author}{Shawe-Taylor, J.}, \bibinfo{author}{Smola, A.J.},
  \bibinfo{author}{Williamson, R.C.}, \bibinfo{year}{2001}.
\newblock \bibinfo{title}{Estimating the support of a high-dimensional
  distribution}.
\newblock \bibinfo{journal}{Neural Computation} \bibinfo{volume}{13},
  \bibinfo{pages}{1443--1471}.
\newblock \DOIprefix\doi{10.1162/089976601750264965}.
%Type = Article
\bibitem[{Selvaraju et~al.(2020)Selvaraju, Cogswell, Das, Vedantam, Parikh and
  Batra}]{selvaraju_grad-cam_2020}
\bibinfo{author}{Selvaraju, R.R.}, \bibinfo{author}{Cogswell, M.},
  \bibinfo{author}{Das, A.}, \bibinfo{author}{Vedantam, R.},
  \bibinfo{author}{Parikh, D.}, \bibinfo{author}{Batra, D.},
  \bibinfo{year}{2020}.
\newblock \bibinfo{title}{Grad-{CAM}: {Visual} {Explanations} from {Deep}
  {Networks} via {Gradient}-{Based} {Localization}}.
\newblock \bibinfo{journal}{International Journal of Computer Vision}
  \bibinfo{volume}{128}, \bibinfo{pages}{336--359}.
\newblock \URLprefix \url{https://doi.org/10.1007/s11263-019-01228-7},
  \DOIprefix\doi{10.1007/s11263-019-01228-7}.
%Type = Inproceedings
\bibitem[{Simarro~Viana et~al.(2021)Simarro~Viana, de~la Rosa, Vande~Vyvere,
  Robben, Sima and Investigators}]{simarro_viana_unsupervised_2021}
\bibinfo{author}{Simarro~Viana, J.}, \bibinfo{author}{de~la Rosa, E.},
  \bibinfo{author}{Vande~Vyvere, T.}, \bibinfo{author}{Robben, D.},
  \bibinfo{author}{Sima, D.M.}, \bibinfo{author}{Investigators, C.T.P.a.},
  \bibinfo{year}{2021}.
\newblock \bibinfo{title}{Unsupervised {3D} {Brain} {Anomaly} {Detection}}, in:
  \bibinfo{editor}{Crimi, A.}, \bibinfo{editor}{Bakas, S.} (Eds.),
  \bibinfo{booktitle}{Brainlesion: {Glioma}, {Multiple} {Sclerosis}, {Stroke}
  and {Traumatic} {Brain} {Injuries}}, \bibinfo{publisher}{Springer
  International Publishing}, \bibinfo{address}{Cham}. pp.
  \bibinfo{pages}{133--142}.
%Type = Inproceedings
\bibitem[{Sohn et~al.(2022)Sohn, Li, Yoon, Jin and
  Pfister}]{sohn_learning_2022}
\bibinfo{author}{Sohn, K.}, \bibinfo{author}{Li, C.L.}, \bibinfo{author}{Yoon,
  J.}, \bibinfo{author}{Jin, M.}, \bibinfo{author}{Pfister, T.},
  \bibinfo{year}{2022}.
\newblock \bibinfo{title}{Learning and {Evaluating} {Representations} for
  {Deep} {One}-{Class} {Classification}}, in: \bibinfo{booktitle}{International
  {Conference} on {Learning} {Representations}}.
\newblock \URLprefix \url{https://openreview.net/forum?id=HCSgyPUfeDj}.
%Type = Inproceedings
\bibitem[{Sun et~al.(2017)Sun, Cachia, Rivière, Fischer, Makin and
  Mangin}]{sun_congenital_2017}
\bibinfo{author}{Sun, Z.Y.}, \bibinfo{author}{Cachia, A.},
  \bibinfo{author}{Rivière, D.}, \bibinfo{author}{Fischer, C.},
  \bibinfo{author}{Makin, T.}, \bibinfo{author}{Mangin, J.F.},
  \bibinfo{year}{2017}.
\newblock \bibinfo{title}{Congenital unilateral upper limb absence flattens the
  contralateral hand knob}, in: \bibinfo{booktitle}{Organisation for {Human}
  {Brain} {Mapping}}, \bibinfo{address}{Vancouver, Canada}.
%Type = Article
\bibitem[{Sun et~al.(2012)Sun, Klöppel, Rivière, Perrot, Frackowiak, Siebner
  and Mangin}]{sun_effect_2012}
\bibinfo{author}{Sun, Z.Y.}, \bibinfo{author}{Klöppel, S.},
  \bibinfo{author}{Rivière, D.}, \bibinfo{author}{Perrot, M.},
  \bibinfo{author}{Frackowiak, R.}, \bibinfo{author}{Siebner, H.},
  \bibinfo{author}{Mangin, J.F.}, \bibinfo{year}{2012}.
\newblock \bibinfo{title}{The effect of handedness on the shape of the central
  sulcus}.
\newblock \bibinfo{journal}{NeuroImage} \bibinfo{volume}{60},
  \bibinfo{pages}{332--339}.
\newblock \URLprefix
  \url{https://www.sciencedirect.com/science/article/pii/S1053811911014522},
  \DOIprefix\doi{10.1016/j.neuroimage.2011.12.050}.
%Type = Article
\bibitem[{Sun et~al.(2009)Sun, Perrot, Tucholka, Rivière and
  Mangin}]{sun_constructing_2009}
\bibinfo{author}{Sun, Z.Y.}, \bibinfo{author}{Perrot, M.},
  \bibinfo{author}{Tucholka, A.}, \bibinfo{author}{Rivière, D.},
  \bibinfo{author}{Mangin, J.F.}, \bibinfo{year}{2009}.
\newblock \bibinfo{title}{Constructing a dictionary of human brain folding
  patterns}.
\newblock \bibinfo{journal}{Medical image computing and computer-assisted
  intervention: MICCAI ... International Conference on Medical Image Computing
  and Computer-Assisted Intervention} \bibinfo{volume}{12},
  \bibinfo{pages}{117--124}.
\newblock \DOIprefix\doi{10.1007/978-3-642-04271-3_15}.
%Type = Article
\bibitem[{Tovar-Moll et~al.(2007)Tovar-Moll, Moll, de~Oliveira-Souza, Bramati,
  Andreiuolo and Lent}]{tovar-moll_neuroplasticity_2007}
\bibinfo{author}{Tovar-Moll, F.}, \bibinfo{author}{Moll, J.},
  \bibinfo{author}{de~Oliveira-Souza, R.}, \bibinfo{author}{Bramati, I.},
  \bibinfo{author}{Andreiuolo, P.A.}, \bibinfo{author}{Lent, R.},
  \bibinfo{year}{2007}.
\newblock \bibinfo{title}{Neuroplasticity in {Human} {Callosal} {Dysgenesis}:
  {A} {Diffusion} {Tensor} {Imaging} {Study}}.
\newblock \bibinfo{journal}{Cerebral Cortex} \bibinfo{volume}{17},
  \bibinfo{pages}{531--541}.
\newblock \URLprefix \url{https://doi.org/10.1093/cercor/bhj178},
  \DOIprefix\doi{10.1093/cercor/bhj178}.
%Type = Inproceedings
\bibitem[{Tschuchnig and Gadermayr(2022)}]{tschuchnig_anomaly_2022}
\bibinfo{author}{Tschuchnig, M.E.}, \bibinfo{author}{Gadermayr, M.},
  \bibinfo{year}{2022}.
\newblock \bibinfo{title}{Anomaly {Detection} in {Medical} {Imaging} - {A}
  {Mini} {Review}}, in: \bibinfo{editor}{Haber, P.},
  \bibinfo{editor}{Lampoltshammer, T.J.}, \bibinfo{editor}{Leopold, H.},
  \bibinfo{editor}{Mayr, M.} (Eds.), \bibinfo{booktitle}{Data {Science} –
  {Analytics} and {Applications}}, \bibinfo{publisher}{Springer Fachmedien},
  \bibinfo{address}{Wiesbaden}. pp. \bibinfo{pages}{33--38}.
\newblock \DOIprefix\doi{10.1007/978-3-658-36295-9_5}.
%Type = Article
\bibitem[{Van~Essen et~al.(2013)Van~Essen, Smith, Barch, Behrens, Yacoub,
  Ugurbil and {WU-Minn HCP Consortium}}]{van_essen_wu-minn_2013}
\bibinfo{author}{Van~Essen, D.C.}, \bibinfo{author}{Smith, S.M.},
  \bibinfo{author}{Barch, D.M.}, \bibinfo{author}{Behrens, T.E.J.},
  \bibinfo{author}{Yacoub, E.}, \bibinfo{author}{Ugurbil, K.},
  \bibinfo{author}{{WU-Minn HCP Consortium}}, \bibinfo{year}{2013}.
\newblock \bibinfo{title}{The {WU}-{Minn} {Human} {Connectome} {Project}: an
  overview}.
\newblock \bibinfo{journal}{NeuroImage} \bibinfo{volume}{80},
  \bibinfo{pages}{62--79}.
\newblock \DOIprefix\doi{10.1016/j.neuroimage.2013.05.041}.
%Type = Article
\bibitem[{de~Vareilles et~al.(2022)de~Vareilles, Rivière, Sun, Fischer, Leroy,
  Neumane, Stopar, Eijsermans, Ballu, Tataranno, Benders, Mangin and
  Dubois}]{de_vareilles_shape_2022}
\bibinfo{author}{de~Vareilles, H.}, \bibinfo{author}{Rivière, D.},
  \bibinfo{author}{Sun, Z.Y.}, \bibinfo{author}{Fischer, C.},
  \bibinfo{author}{Leroy, F.}, \bibinfo{author}{Neumane, S.},
  \bibinfo{author}{Stopar, N.}, \bibinfo{author}{Eijsermans, R.},
  \bibinfo{author}{Ballu, M.}, \bibinfo{author}{Tataranno, M.L.},
  \bibinfo{author}{Benders, M.}, \bibinfo{author}{Mangin, J.F.},
  \bibinfo{author}{Dubois, J.}, \bibinfo{year}{2022}.
\newblock \bibinfo{title}{Shape variability of the central sulcus in the
  developing brain: {A} longitudinal descriptive and predictive study in
  preterm infants}.
\newblock \bibinfo{journal}{NeuroImage} \bibinfo{volume}{251},
  \bibinfo{pages}{118837}.
\newblock \URLprefix
  \url{https://www.sciencedirect.com/science/article/pii/S1053811921011083},
  \DOIprefix\doi{10.1016/j.neuroimage.2021.118837}.
%Type = Article
\bibitem[{Wachinger et~al.(2015)Wachinger, Golland, Kremen, Fischl and
  Reuter}]{wachinger_brainprint_2015}
\bibinfo{author}{Wachinger, C.}, \bibinfo{author}{Golland, P.},
  \bibinfo{author}{Kremen, W.}, \bibinfo{author}{Fischl, B.},
  \bibinfo{author}{Reuter, M.}, \bibinfo{year}{2015}.
\newblock \bibinfo{title}{{BrainPrint}: {A} discriminative characterization of
  brain morphology}.
\newblock \bibinfo{journal}{NeuroImage} \bibinfo{volume}{109},
  \bibinfo{pages}{232--248}.
\newblock \URLprefix
  \url{https://www.sciencedirect.com/science/article/pii/S1053811915000476},
  \DOIprefix\doi{10.1016/j.neuroimage.2015.01.032}.
%Type = Article
\bibitem[{Weiner et~al.(2014)Weiner, Golarai, Caspers, Chuapoco, Mohlberg,
  Zilles, Amunts and Grill-Spector}]{weiner_mid-fusiform_2014}
\bibinfo{author}{Weiner, K.S.}, \bibinfo{author}{Golarai, G.},
  \bibinfo{author}{Caspers, J.}, \bibinfo{author}{Chuapoco, M.R.},
  \bibinfo{author}{Mohlberg, H.}, \bibinfo{author}{Zilles, K.},
  \bibinfo{author}{Amunts, K.}, \bibinfo{author}{Grill-Spector, K.},
  \bibinfo{year}{2014}.
\newblock \bibinfo{title}{The mid-fusiform sulcus: a landmark identifying both
  cytoarchitectonic and functional divisions of human ventral temporal cortex}.
\newblock \bibinfo{journal}{NeuroImage} \bibinfo{volume}{84},
  \bibinfo{pages}{453--465}.
%Type = Article
\bibitem[{Yousry et~al.(1997)Yousry, Schmid, Alkadhi, Schmidt, Peraud, Buettner
  and Winkler}]{yousry_localization_1997}
\bibinfo{author}{Yousry, T.A.}, \bibinfo{author}{Schmid, U.D.},
  \bibinfo{author}{Alkadhi, H.}, \bibinfo{author}{Schmidt, D.},
  \bibinfo{author}{Peraud, A.}, \bibinfo{author}{Buettner, A.},
  \bibinfo{author}{Winkler, P.}, \bibinfo{year}{1997}.
\newblock \bibinfo{title}{Localization of the motor hand area to a knob on the
  precentral gyrus. {A} new landmark.}
\newblock \bibinfo{journal}{Brain} \bibinfo{volume}{120},
  \bibinfo{pages}{141--157}.
\newblock \URLprefix \url{https://doi.org/10.1093/brain/120.1.141},
  \DOIprefix\doi{10.1093/brain/120.1.141}.
%Type = Article
\bibitem[{Yücel et~al.(2003)Yücel, Wood, Phillips, Stuart, Smith, Yung,
  Velakoulis, Mcgorry and Pantelis}]{yucel_morphology_2003}
\bibinfo{author}{Yücel, M.}, \bibinfo{author}{Wood, S.J.},
  \bibinfo{author}{Phillips, L.J.}, \bibinfo{author}{Stuart, G.W.},
  \bibinfo{author}{Smith, D.J.}, \bibinfo{author}{Yung, A.},
  \bibinfo{author}{Velakoulis, D.}, \bibinfo{author}{Mcgorry, P.D.},
  \bibinfo{author}{Pantelis, C.}, \bibinfo{year}{2003}.
\newblock \bibinfo{title}{Morphology of the anterior cingulate cortex in young
  men at ultra-high risk of developing a psychotic illness}.
\newblock \bibinfo{journal}{The British Journal of Psychiatry}
  \bibinfo{volume}{182}, \bibinfo{pages}{518--524}.
\newblock \bibinfo{note}{Publisher: Cambridge University Press}.

\end{thebibliography}
%%Harvard
\bibliographystyle{model2-names.bst}\biboptions{authoryear}
%\bibliography{refs}

\end{document}